\documentclass[%
preprint,
superscriptaddress,
showpacs,
showpacs,preprintnumbers,
amsmath,amssymb,
aps,
]{revtex4-1}

\usepackage[dvipdfmx]{graphicx}
\usepackage{dcolumn}
\usepackage{bm}
\usepackage{multirow}
\usepackage[mathlines]{lineno}
\usepackage{comment}

\usepackage{xr}
\usepackage{color}

\begin{document}

\preprint{APS/123-QED}

\title{Observation of superconductivity and its enhancement at the charge density wave critical point in LaAgSb$_2$}

\author{Kazuto~Akiba}
\email{akb@okayama-u.ac.jp}
\affiliation{
Graduate School of Natural Science and Technology, 
Okayama University, Okayama 700-8530, Japan
}

\author{Nobuaki~Umeshita}
\affiliation{
Graduate School of Natural Science and Technology, 
Okayama University, Okayama 700-8530, Japan
}

\author{Tatsuo~C.~Kobayashi}
\affiliation{
Graduate School of Natural Science and Technology, 
Okayama University, Okayama 700-8530, Japan
}

\date{\today}

\begin{abstract}
We discover superconductivity (SC) in LaAgSb$_2$ at ambient pressure and its close correlation with a charge density wave (CDW) under pressure. 
The superconducting transition temperature ($T_c$) exhibits a sharp peak at the CDW critical pressure of 3.2 GPa. 
We demonstrate that the carriers inhabiting the Sb-square net is crucial not only in the formation of CDW but also in SC for their relatively strong electron-phonon coupling (EPC). 
Furthermore, theoretical EPC strength in pristine LaAgSb$_2$ cannot explain the observed peak with $T_c\sim 1$ K,
which indicates that an additional mechanism reinforces SC only around the CDW critical pressure.
\end{abstract}

\maketitle

Correlation between the superconductivity (SC) and other orders can help identify the mechanism behind SC and offers insight to improve the superconducting transition temperature ($T_c$).
A well-known example is the correlation between SC and magnetism.
When a magnetic phase transition temperature is continuously suppressed toward absolute-zero temperature by an external parameter,
the phase transition can be triggered even at 0 K,
which is known as a quantum critical point (QCP).
Around the QCP,
various physical quantities demonstrate anomalous temperature ($T$) dependence
due to the spin fluctuation \cite{Moriya_SCR}, which is called non-Fermi liquid behavior.
Interestingly, unconventional SCs have been observed near the pressure- ($P$-) induced QCP
in several heavy fermion systems \cite{Movshovich_1996, Grosche_1996, Mathur_1998},
which result in a dome-like superconducting phase in the $T$--$P$ phase diagram.
This suggests that the spin fluctuation plays an important role on the pairing mechanism in this material class.

Alternatively, 
the correlation between SC and charge density wave (CDW) has also attracted considerable attention.
In several representative materials,
the emergence of SC or substantial enhancement of $T_c$
occurs near the CDW critical point \cite{Morosan_2005, Kusmartseva_2009, Gruner_2017, Chen_2021, Fu_2021},
which is often associated with a non-monotonic change of power law in the $\rho$--$T$ curve at low temperatures \cite{Kusmartseva_2009, Gruner_2017}.
Intriguingly, these phenomena have several similar features as those of the magnetic QCP case, which
suggests that these two phenomena have a possible commonality.
However, a lack of model materials has hindered the elucidation of the essential relationship between SC and CDW.

Herein, we focus on the layered intermetallic compound LaAgSb$_2$ as an ideal platform to achieve the aforementioned goals.
At ambient pressure,
LaAgSb$_2$ exhibits sequential CDW transitions at $T_{CDW1}=210$ K and $T_{CDW2}=190$ K \cite{Myers_1999a, Song_2003},
whose critical pressures are determined as $P_{CDW1}\sim 3.2 $ GPa and $P_{CDW2}\sim 1.7 $ GPa,
respectively \cite{Akiba_2021}.
Although the origin of CDW2 is not conclusive at present,
CDW1 is assumed to be derived from the nesting within the characteristic hollow-shaped Fermi surface (FS) \cite{Song_2003, Bosak_2021, Akiba_2022}.
Previous studies have reported that
the hollow-shaped FS exhibits Dirac-like linear dispersion at the Fermi level \cite{Wang_2012Oct, Shi_2016}.
The emergence of the linear band crossing in Sb- or Bi-square-net materials
can be understood
as a result of band folding associated with the $4^4$-net structure \cite{Hoffmann_1987, Klemenz_2019, Klemenz_2020}.
Intriguingly, isostructural compounds LaAuSb$_2$ \cite{Du_2020} and LaCuSb$_2$ \cite{Muro_1997} show superconductivity
at 0.6 K and 0.9 K respectively,
the former of which coexists with CDW order.
However, there has been no report on SC in LaAgSb$_2$ thus far.

In this Letter, we reported the discovery of bulk SC in LaAgSb$_2$ at ambient pressure with $T_c\sim 0.3$ K.
SC exhibited intimate correlation with CDW1 under high pressure,
and $T_c$ was significantly enhanced up to $T_c\sim 1$ K only around $P_{CDW1}$.
The theoretical calculation demonstrated that
the Sb-square net is crucial not only as a host of Dirac fermions and CDW1, but also as a primary superconducting layer.
In addition, theoretical $T_c$ of pristine LaAgSb$_2$ cannot reproduce the peak structure under pressure,
which suggests that an additional reinforcement mechanism of SC activated only around the CDW critical point exists.

Single crystal of LaAgSb$_2$ were synthesized using the Sb self-flux method \cite{Myers_1999a, Akiba_2021}.
Temperatures as low as 50 mK were realized using a homemade $^3$He/$^4$He dilution refrigerator.
High pressure was generated using an indenter-type pressure cell (IPC, $P < 4$ GPa) \cite{Kobayashi_2007} and opposed-anvil-type pressure cell (OAPC, $P < 7$ GPa) \cite{Kitagawa_2010}.
Daphne oil 7474 \cite{Murata_2008} was used as a pressure medium.
The pressure in the sample space was determined based on the $T_c$ of Pb set near the sample.
The resistivity measurements were performed using a Model 370 AC resistance bridge (Lake Shore Cryotronics, Inc.) following a standard four-terminal method.
Magnetization measurements were performed using a DC superconducting quantum interference device (Tristan Technologies, Inc.).
In the magnetization measurement, we did not intentionally apply an external magnetic field.
The measurements were done in residual geomagnetic field.
Specific heat was measured by the relaxation method.
The details of the magnetization and specific heat measurements are described in the Supplemental Material \cite{SM_URL}.
For resistivity and magnetization measurements,
the sample was shaped into rectangle with typical dimensions of $1\times 0.5 \times 0.1$ mm$^3$.
For specific heat measurements, relatively large as-grown crystal (typically several tens of milligram) was used.
First-principles calculations were performed by Quantum ESPRESSO package \cite{Giannozzi_2009, Giannozzi_2017, Kawamura_2014}
using scalar-relativistic ultrasoft pseudopotentials.
Electron-phonon coupling (EPC) and $T_c$ were calculated
using Wannier90 \cite{Pizzi_2020} and EPW \cite{Ponce_2016} codes.
The visualization of orbital projections on FS was performed by FermiSurfer \cite{Kawamura_2019}.
 Calculation and verification of the results are described in detail in the Supplemental Material \cite{SM_URL}.

\begin{figure}[]
\centering
\includegraphics[]{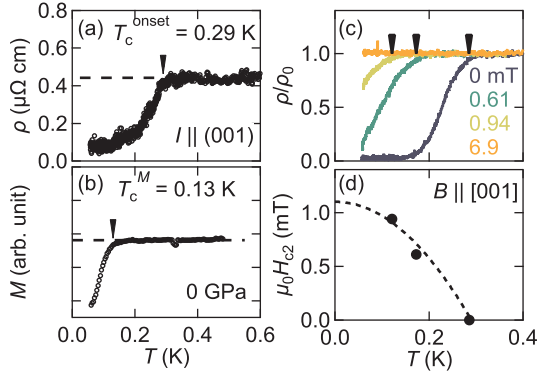}
\caption{
(a) Temperature dependence of the in-plane resistivity $\rho$ at ambient pressure.
The onset temperature of the superconducting transition is determined as
$T_c^{onset}=0.29$ K.
(b) Temperature dependence of the magnetization $M$ at ambient pressure.
The onset of the magnetization anomaly is determined as $T_c^{M}=0.13$ K.
(c)  Temperature dependence of $\rho$ normalized by the value at 0.4 K ($\rho_0$) under several external magnetic fields. Solid arrows indicate the onset of superconducting transition.
(d) Temperature dependence of the upper critical field $\mu_0 H_{c2}$ obtained from the data shown in (c).
The dashed line represents $\mu_0 H_{c2}$ (mT)$=1.1[1-(T/0.29)^2]$.
\label{fig01}}
\end{figure}

Figure \ref{fig01}(a) shows the temperature dependence of the in-plane resistivity $\rho$ [current within the (001) plane] at ambient pressure.
The residual resistivity ratio deduced from $\rho$ at 300 K and 0.6 K was $\sim 88$,
and $T_{CDW1}=210$ K was clearly observed (not shown).
As depicted in Fig. \ref{fig01}(a),
an abrupt decrease in $\rho$ was observed,
whose onset temperature can be defined as $T_c^{onset}=0.29$ K, indicated by the arrow.
This observation strongly suggests that the superconducting state exists at ambient pressure.

Figure \ref{fig01}(b) plots the temperature dependence of the magnetization $M$ for identical sample,
which shows distinct anomaly at $T_c^{M}=0.13$ K.
In addition, the superconducting volume fraction was estimated as $\sim 39$ \% at 60 mK,
as described in Supplemental Material \cite{SM_URL}.
Since the present measurement corresponds to field-cooling under residual geomagnetic field,
the estimated volume fraction can be regarded as lower limit.
The incomplete shielding and apparent difference between $T_c^{M}$ and $T_c^{onset}$
are presumably due to imperfect zero-resistivity even at the lowest $T$ and broad superconducting transition in this sample.

Figure \ref{fig01}(c) shows the temperature dependence of $\rho$
normalized by the value at 0.4 K ($\rho_0$)
under external magnetic fields along [001] ($c$) axis.
We note that there is a sample dependence in the sharpness of the superconducting transition;
in a sample investigated in Fig. \ref{fig01}(c),
we can recognize more clear zero resistivity than Fig. \ref{fig01}(a).
The onset temperature of superconducting transition is systematically decreases as the magnetic field increases,
and SC is completely absent under 6.9 mT.
We can estimate from Fig. \ref{fig01}(d) the upper critical field ($\mu_0 H_{c2}$) at 0 K as $\sim 1.1$ mT.

\begin{figure}[]
\centering
\includegraphics[]{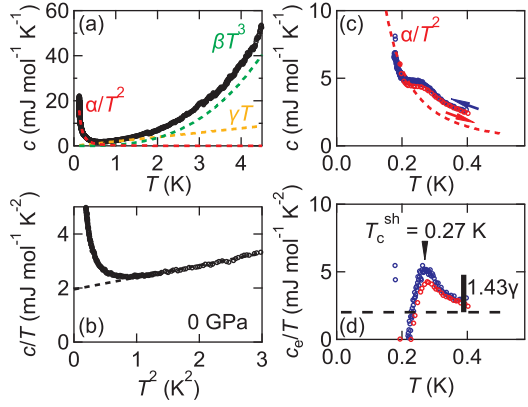}
\caption{
(a) Temperature dependence of the specific heat $c$ at ambient pressure.
Dotted curves represent the individual contributions from nuclei (red), phonons (green), and electrons (orange).
(b) $c/T$ as a function of $T^2$.
The straight line represents the linear fit used for deducing the electron and phonon specific heat coefficients.
(c) Specific heat anomaly below 0.5 K.
Red and blue markers represent the heating and cooling processes, respectively.
Dashed curve represents the nuclear contribution which is proportional to $1/T^2$.
(d) $c_e/T$ as a function of $T$, where $c_e$ represents the electronic specific heat.
Vertical scale represents the expected jump in BCS theory ($\sim 2.8$ mJ mol$^{-1}$ K$^{-2}$).
\label{fig02}}
\end{figure}

To obtain thermodynamic evidence, specific heat at ambient pressure was measured.
Fig. \ref{fig02}(a) shows the temperature dependence of the specific heat $c$.
$c$ is represented as $c=\alpha/T^2+\beta T^3+\gamma T$,
where the first, second, and third term represents the nuclear, phonon, and electron contribution, respectively.
At relatively high temperature above 1 K, the nuclear contribution is negligibly small;
thus, $c$ is dominated by electron and phonon contributions.
This results in the linear dependence in $c/T$--$T^2$ plot,
as shown in Fig. \ref{fig02}(b).
From the slope and vertical intercept of the straight line, $\beta=(444\pm 7) \mu$J mol$^{-1}$ K$^{-4}$ and $\gamma=(1.95\pm 0.01)$mJ mol$^{-1}$ K$^{-2}$ were obtained.
From $\beta$, the Debye temperature was estimated as $\Theta_D=(48R\pi^4/5\beta)^{1/3}=260$ K,
where $R$ represents the gas constant.
The obtained $\gamma$ and $\Theta_D$ are comparable
to values reported in previous experiments \cite{Muro_1997, Inada_2002}.
In contrast, at temperatures below 1 K,
a drastic increase in nuclear contribution was observed, as shown in Fig. \ref{fig02}(a),
which is well scaled by $\alpha/T^2$ with $\alpha=(259.4\pm 0.5) \mu$J mol$^{-1}$ K.
By first-principles calculation, we confirmed notably large nuclear quadropole resonance frequency of Sb
due to large electromagnetic field gradient around Sb site,
which is ascribed to be the origin of Schottky-type behavior below 1 K.
Although the anomaly due to the superconducting transition
is quite smaller than nuclear specific heat,
a hump-like anomaly can be discerned at $T_c^{sh}=0.27$ K, as shown in Fig. \ref{fig02}(c).
Figure \ref{fig02}(d) shows $c_e/T$ as a function of $T$,
where the electronic specific heat $c_e$ is obtained by subtracting the dashed curve in Fig. \ref{fig02}(c) from $c$.
Magnitude of the anomaly shows reasonable agreement with
$1.43 \gamma$ $\sim 2.8$ mJ mol$^{-1}$ K$^{-2}$ [black vertical scale in Fig. \ref{fig02}(d)],
which is expected from Bardeen-Cooper-Schrieffer (BCS) theory.
This serves firm evidence for bulk SC at ambient pressure.

\begin{figure}[]
\centering
\includegraphics[]{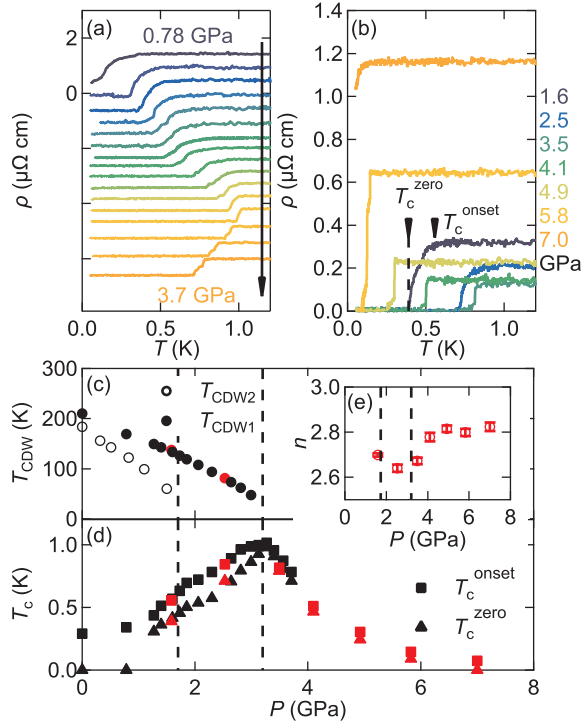}
\caption{
Temperature dependence of $\rho$ (a) up to 3.7 GPa measured using the indenter-type pressure cell (IPC) and
(b) up to 7.0 GPa measured using the opposed-anvil-type pressure cell (OAPC).
Each trace in (a) is vertically shifted for clarity.
The definition of $T_c^{onset}$ and $T_c^{zero}$ are indicated by arrows
in the case of 1.6 GPa in (b).
Pressure dependence of (c) CDW transition and (d) superconducting transition temperatures.
The black and red markers represent the data obtained using IPC and OAPC, respectively.
(e) Pressure dependence of power $n$ in $\rho = \rho_0+A' T^n$ over the range 2.5 K $<T<$ 35 K.
\label{fig03}}
\end{figure}

From the above results,
the coexistence of the CDWs and SC in LaAgSb$_2$ at ambient pressure was established.
Subsequently, their correlation under pressure was investigated.

Figure \ref{fig03}(a) and (b) show the temperature dependence of $\rho$ up to 3.7 GPa
and up to 7.0 GPa, obtained using IPC and OAPC, respectively.
The pressure dependence of CDW and superconducting transition temperatures are summarized in Figs. \ref{fig03}(c) and \ref{fig03}(d), respectively.
The consistency of the data obtained by IPC and OAPC is satisfactory.
Herein, $T_{CDW1}$ was determined from the data presented in Supplemental Material \cite{SM_URL},
and $T_{CDW2}$ was taken from our previous study \cite{Akiba_2021}.
The definitions of $T_{c}^{onset}$ and $T_{c}^{zero}$ are indicated in Fig. \ref{fig03}(b) by arrows.
Although the transition was broad at pressures below 1 GPa,
zero-resistivity above 1.3 GPa was distinctly observed.
Both $T_{c}^{onset}$ and $T_{c}^{zero}$ gradually increased with pressure
and reached their maxima almost exactly at $P_{CDW1}=3.2$ GPa.
Maximum onset temperature of $T_{c}^{onset}=0.98$ K was found to be 3.4 times higher than that measured at ambient pressure.
On the other hand,
we can find no discernible anomaly at $P_{CDW2}\sim 1.7 $ GPa.
Above $P_{CDW1}$,
$T_{c}$ immediately got suppressed,
and its decrement rate was steeper than the increment below $P_{CDW1}$.
This resulted in a distinct peak structure in the $T$--$P$ phase diagram, as shown in Fig. \ref{fig03}(d).
At the highest pressure of 7.0 GPa,
$T_{c}^{onset} = 75$ mK was 1/13 compared to its maximum value,
and $T_{c}^{zero}$ was again below the lowest limit of the present study. 
The obvious correlation with CDW1 further confirms the bulk SC.

\begin{figure}[]
\centering
\includegraphics[]{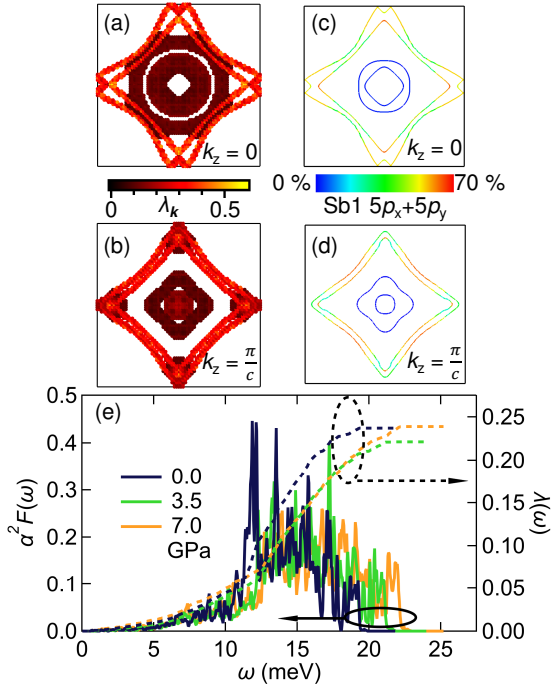}
\caption{
Momentum-resolved electron--phonon coupling strength $\lambda_{\bm{k}}$ on (a) $k_z=0$ and (b) $k_z=\pi/c$ planes at ambient pressure.
$\lambda_{\bm{k}}$ is calculated for $k$ points within the range of $\pm 0.2$ eV from the Fermi level.
The projection of $5p_x+5p_y$ orbital character for Sb1 on (c) $k_z=0$ and (d) $k_z=\pi/c$ planes at ambient pressure.
(e) Eliashberg spectral function $\alpha^2F(\omega)$ (left axis)
and integrated electron--phonon coupling strength $\lambda(\omega)$ (right axis) at 0, 3.5, and 7.0 GPa.
\label{fig04}}
\end{figure}

Here, we discuss the correlation between CDW1 and SC.
Because LaAgSb$_2$ is a non-magnetic material,
the pairing mechanism is primarily assumed to be the EPC.
Thus, we focused on the EPC strength in the momentum space at ambient pressure.
In the following calculations, lattice modulation by CDWs was not considered,
i.e., the results in pristine crystal structures without CDWs.

Figs. \ref{fig04} (a) and (b) represent the distribution of the EPC strength $\lambda_{\bm{k}}$ on $k_z=0$ and $k_z=\pi/c$ planes, respectively.
Carriers with relatively strong EPC ($\lambda_{\bm{k}}\sim 0.3$) are concentrated on the two-dimensional hollow-like surface and elongated pocket at X point,
whereas the rest of the FS exhibited weaker EPC ($\lambda_{\bm{k}}=$0.1--0.15).
As shown in Supplemental Material \cite{SM_URL},
this situation also holds under pressure.
Because the hollow-like FS is the host of the CDW1 nesting,
the abovementioned results lead to the interpretation that the CDW1 and SC compete on identical FS,
which results in their remarkable correlation.

Figs. \ref{fig04} (c) and (d) show the orbital characters of the FS on the $k_z$ planes described in Figs. \ref{fig04} (a) and (b).
Notably the FSs with strong EPC are dominated by the intense $5p_{x,y}$ character of Sb1,
which is a constituent of Sb-square net.
The population of other orbitals were relatively minor, as presented in Supplemental Material \cite{SM_URL}.
These results indicate that the Sb-square net
is responsible not only for the emergence of linear band crossing and nesting of CDW1
but also serves as a primary superconducting layer in this material.
As we previously mentioned, no discernible anomaly of $T_c$ at $P_{CDW2}\sim 1.7 $ GPa was observed.
Given that the SC is primarily responsible for the Sb-$p_{x,y}$,
this behavior can be reasonably understood.
The lattice modulation accompanied by CDW2 is known to be along the $c$ axis \cite{Song_2003}, which is perpendicular to the orbital spread;
thus, a rather minor effect on SC is expected.
In contrast, CDW1 causes lattice modulation along the $a$ axis \cite{Song_2003} within the Sb-square net,
which can directly affect the superconducting properties.

\begin{table}
\caption{\label{tab_tmad}
Theoretical EPC strength ($\lambda$), experimental EPC strength ($\lambda^{exp}$), logarithmic average of phonon frequency ($\omega_{log}$, converted into temperature), and superconducting transition temperature estimated by McMillan--Allen--Dynes formula ($T_c^{MAD}$)
with the Coulomb pseudopotential $\mu_c^*=0.1$.
}
\begin{ruledtabular}
\begin{tabular}{lrrrr}
$P$ (GPa) & $\lambda$ & $\lambda^{exp}$ & $\omega_{log}$ (K) & $T_c^{MAD}$ (mK) \\
\hline
0.0 & 0.237 & 0.34 & 137.4 & 3.1\\
3.5 & 0.221 & 0.40 & 149.0 & 0.92\\
7.0 & 0.239 & 0.29 & 147.0 & 3.8
\end{tabular}
\end{ruledtabular}
\end{table}

Here, we extend our quantitative discussion on the superconducting properties based on McMillan--Allen--Dynes formalism \cite{McMillan_1968, Dynes_1972, Allen_1975}.
Within this framework, the superconducting transition temperature $T_c^{MAD}$ is represented as
\begin{equation}
T_c^{MAD} = \dfrac{\omega_{log}}{1.2}\exp \left(-\dfrac{1.04(1+\lambda)}{\lambda-\mu_c^*(1+0.62\lambda)}\right),
\label{eq_tmad}
\end{equation}
where, $\lambda=\lambda(\omega_{max})$ is the integrated EPC strength defined by
\begin{equation}
\lambda(\omega) = 2\int_0^{\omega} d\omega' \dfrac{\alpha^2F(\omega')}{\omega'},
\label{eq_lambda_int}
\end{equation}
and $\omega_{log}$ is the logarithmic average of the phonon frequency defined as
\begin{equation}
\omega_{log}=\exp\left( \dfrac{2}{\lambda}\int_0^{\omega_{max}} d\omega \ln\omega \dfrac{\alpha^2F(\omega)}{\omega}\right),
\label{eq_omegalog_int}
\end{equation}
where $\omega_{max}$ denotes the maximum of the phonon frequency.
$\alpha^2F(\omega)$ is the Eliashberg spectral function, which can be obtained from first-principles results,
as described in Supplemental Material \cite{SM_URL}.
The calculated $\alpha^2F(\omega)$ at 0, 3.5, and 7.0 GPa are presented in Fig. \ref{fig04}(e)
with corresponding $\lambda(\omega)$ values;
the resulting
$\lambda$, $\omega_{log}$, and $T_c^{MAD}$
are listed in Table \ref{tab_tmad}.
To calculate $T_c^{MAD}$, we assumed a typical Coulomb pseudopotential of $\mu_c^*=0.1$ for metals \cite{Morel_1962}.
The obtained $\lambda$ did not exhibit any notable pressure dependence. Accordingly, $T_c^{MAD}$ was in the order of 1 mK at all pressures,
which is three orders of magnitude smaller than the experimental maximum $T_c \sim 1$ K.
Furthermore, theoretical upper limit of $T_c$ could be calculated by setting $\mu_c^*=0$, i.e., by
completely disregarding the Coulomb interaction.
The averaged upper limit was $\sim0.49$ K, which is lower than the experimental maximum. 

Here, we can deduce experimental EPC strength $\lambda^{exp}$ based on \cite{McMillan_1968}
\begin{equation}
\lambda^{exp}=\dfrac{1.04+\mu_c^* \ln(\Theta_D/1.45T_c)}{(1-0.62\mu_c^*)\ln(\Theta_D/1.45T_c) -1.04}.
\end{equation}
Assuming $\Theta_D=260$ K (estimated at ambient pressure), $\mu_c^*=0.1$, and experimental $T_c^{onset}$,
calculated $\lambda^{exp}$ values are listed in Table \ref{tab_tmad}.
At 7.0 GPa, far from $P_{CDW1}$, the difference between theoretical and experimental $\lambda$ is small compared with other pressures, which
supports the weak coupling nature of pristine LaAgSb$_2$.
In contrast, near the $P_{CDW1}$, a $\lambda$ nearly two times larger than the theoretical value is necessary
to realize the observed maximum $T_c$ assuming the EPC mechanism.

Consequently, the above results lead to a conclusion that there exists additional mechanism that significantly reinforces the SC only around $P_{CDW1}$.
In the real case, the CDW and superconducting gaps should compete for density of states on the FS below $P_{CDW1}$ \cite{Bilbro_1976}.
Although this conventional picture may partially account for the increasing trend of $T_c$ below $P_{CDW1}$, it cannot explain
significant difference in $T_c$ between experiment and theory at $P_{CDW1}$
and rapid decrease above $P_{CDW1}$.

Since above calculation does not explicitly consider the structural modification in the CDW states,
we do not rule out an EPC origin, possibly related with a phonon softening by CDW1 formation \cite{Chen_2017, Bosak_2021},
at the present stage.
Another possible mechanism is of electronic origin, i.e., a charge fluctuation around $P_{CDW1}$.
Assuming a spin-singlet SC, the effective interaction of the Cooper pairs $V^s(\bm{k}, \bm{k}')$ is given by
$V^s(\bm{k}, \bm{k}')=U+3U^2\chi_s(\bm{k}-\bm{k}')/2-U^2\chi_c(\bm{k}-\bm{k}')/2$ \cite{Yanase_2003},
where $U$, $\chi_s(\bm{k}-\bm{k}')$, and $\chi_c(\bm{k}-\bm{k}')$ represent Coulomb repulsion, spin susceptibility, and charge susceptibility, respectively.
Since the increase of $\chi_c$ brings negative contribution on $V^s$, it can enhance the pairing attraction of the spin-singlet SC.
Due to the coexistence of CDW1 and SC on identical FS and the absence of magnetism, the effect of charge fluctuation may stand out in the present case.
We also note that a non-monotonic change in the exponent $n$ in $\rho=\rho_0+A'T^n$ was observed around 4 GPa, as shown in Fig. \ref{fig03}(e),
which have been discussed in a context of possible quantum criticality at a CDW critical point \cite{Kusmartseva_2009, Gruner_2017}.
To distinguish above possibilities,
further rigorous considerations on EPC and the role of charge fluctuation on SC around $P_{CDW1}$ must be addressed,
which is beyond the scope of the present study.
The detailed mechanism of the reinforcement of SC is left as an open question for future study.

In conclusion, we observed the coexisting SC and CDW states in LaAgSb$_2$ at ambient pressure and their close correlation under high pressure.
The EPC calculations revealed that
the Sb-square net plays a vital role not only in the formation of linear band crossing and CDW1 but also in the emergence of SC.
Our results highlight the strong reinforcement of SC activated only around the CDW1 critical point.
LaAgSb$_2$ under high pressure serves an ideal platform
to elucidate the essence of the correlation between SC and CDW
and to seek a route to improve $T_c$ based on critical points of charge orders.
\begin{acknowledgments}
We thank J. Otsuki 
and H. Harima
for valuable discussion and comments.
This research was supported by JSPS KAKENHI Grants No. 19K14660, 21H01042, and 22K14006.
\end{acknowledgments}

\bibliography{reference}

\begin{thebibliography}{39}%
\makeatletter
\providecommand \@ifxundefined [1]{%
 \@ifx{#1\undefined}
}%
\providecommand \@ifnum [1]{%
 \ifnum #1\expandafter \@firstoftwo
 \else \expandafter \@secondoftwo
 \fi
}%
\providecommand \@ifx [1]{%
 \ifx #1\expandafter \@firstoftwo
 \else \expandafter \@secondoftwo
 \fi
}%
\providecommand \natexlab [1]{#1}%
\providecommand \enquote  [1]{``#1''}%
\providecommand \bibnamefont  [1]{#1}%
\providecommand \bibfnamefont [1]{#1}%
\providecommand \citenamefont [1]{#1}%
\providecommand \href@noop [0]{\@secondoftwo}%
\providecommand \href [0]{\begingroup \@sanitize@url \@href}%
\providecommand \@href[1]{\@@startlink{#1}\@@href}%
\providecommand \@@href[1]{\endgroup#1\@@endlink}%
\providecommand \@sanitize@url [0]{\catcode `\\12\catcode `\$12\catcode
  `\&12\catcode `\#12\catcode `\^12\catcode `\_12\catcode `\%12\relax}%
\providecommand \@@startlink[1]{}%
\providecommand \@@endlink[0]{}%
\providecommand \url  [0]{\begingroup\@sanitize@url \@url }%
\providecommand \@url [1]{\endgroup\@href {#1}{\urlprefix }}%
\providecommand \urlprefix  [0]{URL }%
\providecommand \Eprint [0]{\href }%
\providecommand \doibase [0]{http://dx.doi.org/}%
\providecommand \selectlanguage [0]{\@gobble}%
\providecommand \bibinfo  [0]{\@secondoftwo}%
\providecommand \bibfield  [0]{\@secondoftwo}%
\providecommand \translation [1]{[#1]}%
\providecommand \BibitemOpen [0]{}%
\providecommand \bibitemStop [0]{}%
\providecommand \bibitemNoStop [0]{.\EOS\space}%
\providecommand \EOS [0]{\spacefactor3000\relax}%
\providecommand \BibitemShut  [1]{\csname bibitem#1\endcsname}%
\let\auto@bib@innerbib\@empty
\bibitem [{\citenamefont {Moriya}(1985)}]{Moriya_SCR}%
  \BibitemOpen
  \bibfield  {author} {\bibinfo {author} {\bibfnamefont {T.}~\bibnamefont
  {Moriya}},\ }\href@noop {} {\emph {\bibinfo {title} {Spin Fluctuation in
  Itinerant Electron Magnetism}}}\ (\bibinfo  {publisher} {Springer, Berlin},\
  \bibinfo {year} {1985})\BibitemShut {NoStop}%
\bibitem [{\citenamefont {Movshovich}\ \emph {et~al.}(1996)\citenamefont
  {Movshovich}, \citenamefont {Graf}, \citenamefont {Mandrus}, \citenamefont
  {Thompson}, \citenamefont {Smith},\ and\ \citenamefont
  {Fisk}}]{Movshovich_1996}%
  \BibitemOpen
  \bibfield  {author} {\bibinfo {author} {\bibfnamefont {R.}~\bibnamefont
  {Movshovich}}, \bibinfo {author} {\bibfnamefont {T.}~\bibnamefont {Graf}},
  \bibinfo {author} {\bibfnamefont {D.}~\bibnamefont {Mandrus}}, \bibinfo
  {author} {\bibfnamefont {J.~D.}\ \bibnamefont {Thompson}}, \bibinfo {author}
  {\bibfnamefont {J.~L.}\ \bibnamefont {Smith}}, \ and\ \bibinfo {author}
  {\bibfnamefont {Z.}~\bibnamefont {Fisk}},\ }\href@noop {} {\bibfield
  {journal} {\bibinfo  {journal} {Phys. Rev. B}\ }\textbf {\bibinfo {volume}
  {53}},\ \bibinfo {pages} {8241} (\bibinfo {year} {1996})}\BibitemShut
  {NoStop}%
\bibitem [{\citenamefont {Grosche}\ \emph {et~al.}(1996)\citenamefont
  {Grosche}, \citenamefont {Julian}, \citenamefont {Mathur},\ and\
  \citenamefont {Lonzarich}}]{Grosche_1996}%
  \BibitemOpen
  \bibfield  {author} {\bibinfo {author} {\bibfnamefont {F.}~\bibnamefont
  {Grosche}}, \bibinfo {author} {\bibfnamefont {S.}~\bibnamefont {Julian}},
  \bibinfo {author} {\bibfnamefont {N.}~\bibnamefont {Mathur}}, \ and\ \bibinfo
  {author} {\bibfnamefont {G.}~\bibnamefont {Lonzarich}},\ }\href@noop {}
  {\bibfield  {journal} {\bibinfo  {journal} {Physica B: Condensed Matter}\
  }\textbf {\bibinfo {volume} {223-224}},\ \bibinfo {pages} {50} (\bibinfo
  {year} {1996})}\BibitemShut {NoStop}%
\bibitem [{\citenamefont {Mathur}\ \emph {et~al.}(1998)\citenamefont {Mathur},
  \citenamefont {Grosche}, \citenamefont {Julian}, \citenamefont {Walker},
  \citenamefont {Freye}, \citenamefont {Haselwimmer},\ and\ \citenamefont
  {Lonzarich}}]{Mathur_1998}%
  \BibitemOpen
  \bibfield  {author} {\bibinfo {author} {\bibfnamefont {N.~D.}\ \bibnamefont
  {Mathur}}, \bibinfo {author} {\bibfnamefont {F.~M.}\ \bibnamefont {Grosche}},
  \bibinfo {author} {\bibfnamefont {S.~R.}\ \bibnamefont {Julian}}, \bibinfo
  {author} {\bibfnamefont {I.~R.}\ \bibnamefont {Walker}}, \bibinfo {author}
  {\bibfnamefont {D.~M.}\ \bibnamefont {Freye}}, \bibinfo {author}
  {\bibfnamefont {R.~K.~W.}\ \bibnamefont {Haselwimmer}}, \ and\ \bibinfo
  {author} {\bibfnamefont {G.~G.}\ \bibnamefont {Lonzarich}},\ }\href@noop {}
  {\bibfield  {journal} {\bibinfo  {journal} {Nature (London)}\ }\textbf
  {\bibinfo {volume} {394}},\ \bibinfo {pages} {39} (\bibinfo {year}
  {1998})}\BibitemShut {NoStop}%
\bibitem [{\citenamefont {Morosan}\ \emph {et~al.}(2006)\citenamefont
  {Morosan}, \citenamefont {Zandbergen}, \citenamefont {Dennis}, \citenamefont
  {Bos}, \citenamefont {Onose}, \citenamefont {Klimczuk}, \citenamefont
  {Ramirez}, \citenamefont {Ong},\ and\ \citenamefont {Cava}}]{Morosan_2005}%
  \BibitemOpen
  \bibfield  {author} {\bibinfo {author} {\bibfnamefont {E.}~\bibnamefont
  {Morosan}}, \bibinfo {author} {\bibfnamefont {H.~W.}\ \bibnamefont
  {Zandbergen}}, \bibinfo {author} {\bibfnamefont {B.~S.}\ \bibnamefont
  {Dennis}}, \bibinfo {author} {\bibfnamefont {J.~W.~G.}\ \bibnamefont {Bos}},
  \bibinfo {author} {\bibfnamefont {Y.}~\bibnamefont {Onose}}, \bibinfo
  {author} {\bibfnamefont {T.}~\bibnamefont {Klimczuk}}, \bibinfo {author}
  {\bibfnamefont {A.~P.}\ \bibnamefont {Ramirez}}, \bibinfo {author}
  {\bibfnamefont {N.~P.}\ \bibnamefont {Ong}}, \ and\ \bibinfo {author}
  {\bibfnamefont {R.~J.}\ \bibnamefont {Cava}},\ }\href@noop {} {\bibfield
  {journal} {\bibinfo  {journal} {Nat. Phys}\ }\textbf {\bibinfo {volume}
  {2}},\ \bibinfo {pages} {544} (\bibinfo {year} {2006})}\BibitemShut {NoStop}%
\bibitem [{\citenamefont {Kusmartseva}\ \emph {et~al.}(2009)\citenamefont
  {Kusmartseva}, \citenamefont {Sipos}, \citenamefont {Berger}, \citenamefont
  {Forr\'o},\ and\ \citenamefont {Tuti\ifmmode~\check{s}\else
  \v{s}\fi{}}}]{Kusmartseva_2009}%
  \BibitemOpen
  \bibfield  {author} {\bibinfo {author} {\bibfnamefont {A.~F.}\ \bibnamefont
  {Kusmartseva}}, \bibinfo {author} {\bibfnamefont {B.}~\bibnamefont {Sipos}},
  \bibinfo {author} {\bibfnamefont {H.}~\bibnamefont {Berger}}, \bibinfo
  {author} {\bibfnamefont {L.}~\bibnamefont {Forr\'o}}, \ and\ \bibinfo
  {author} {\bibfnamefont {E.}~\bibnamefont {Tuti\ifmmode~\check{s}\else
  \v{s}\fi{}}},\ }\href@noop {} {\bibfield  {journal} {\bibinfo  {journal}
  {Phys. Rev. Lett.}\ }\textbf {\bibinfo {volume} {103}},\ \bibinfo {pages}
  {236401} (\bibinfo {year} {2009})}\BibitemShut {NoStop}%
\bibitem [{\citenamefont {Gruner}\ \emph {et~al.}(2017)\citenamefont {Gruner},
  \citenamefont {Jang}, \citenamefont {Huesges}, \citenamefont {Cardoso-Gil},
  \citenamefont {Fecher}, \citenamefont {Koza}, \citenamefont {Stockert},
  \citenamefont {Mackenzie}, \citenamefont {Brando},\ and\ \citenamefont
  {Geibel}}]{Gruner_2017}%
  \BibitemOpen
  \bibfield  {author} {\bibinfo {author} {\bibfnamefont {T.}~\bibnamefont
  {Gruner}}, \bibinfo {author} {\bibfnamefont {D.}~\bibnamefont {Jang}},
  \bibinfo {author} {\bibfnamefont {Z.}~\bibnamefont {Huesges}}, \bibinfo
  {author} {\bibfnamefont {R.}~\bibnamefont {Cardoso-Gil}}, \bibinfo {author}
  {\bibfnamefont {G.~H.}\ \bibnamefont {Fecher}}, \bibinfo {author}
  {\bibfnamefont {M.~M.}\ \bibnamefont {Koza}}, \bibinfo {author}
  {\bibfnamefont {O.}~\bibnamefont {Stockert}}, \bibinfo {author}
  {\bibfnamefont {A.~P.}\ \bibnamefont {Mackenzie}}, \bibinfo {author}
  {\bibfnamefont {M.}~\bibnamefont {Brando}}, \ and\ \bibinfo {author}
  {\bibfnamefont {C.}~\bibnamefont {Geibel}},\ }\href@noop {} {\bibfield
  {journal} {\bibinfo  {journal} {Nat. Phys.}\ }\textbf {\bibinfo {volume}
  {13}},\ \bibinfo {pages} {967} (\bibinfo {year} {2017})}\BibitemShut
  {NoStop}%
\bibitem [{\citenamefont {Chen}\ \emph {et~al.}(2021)\citenamefont {Chen},
  \citenamefont {Wang}, \citenamefont {Yin}, \citenamefont {Gu}, \citenamefont
  {Jiang}, \citenamefont {Tu}, \citenamefont {Gong}, \citenamefont {Uwatoko},
  \citenamefont {Sun}, \citenamefont {Lei}, \citenamefont {Hu},\ and\
  \citenamefont {Cheng}}]{Chen_2021}%
  \BibitemOpen
  \bibfield  {author} {\bibinfo {author} {\bibfnamefont {K.~Y.}\ \bibnamefont
  {Chen}}, \bibinfo {author} {\bibfnamefont {N.~N.}\ \bibnamefont {Wang}},
  \bibinfo {author} {\bibfnamefont {Q.~W.}\ \bibnamefont {Yin}}, \bibinfo
  {author} {\bibfnamefont {Y.~H.}\ \bibnamefont {Gu}}, \bibinfo {author}
  {\bibfnamefont {K.}~\bibnamefont {Jiang}}, \bibinfo {author} {\bibfnamefont
  {Z.~J.}\ \bibnamefont {Tu}}, \bibinfo {author} {\bibfnamefont {C.~S.}\
  \bibnamefont {Gong}}, \bibinfo {author} {\bibfnamefont {Y.}~\bibnamefont
  {Uwatoko}}, \bibinfo {author} {\bibfnamefont {J.~P.}\ \bibnamefont {Sun}},
  \bibinfo {author} {\bibfnamefont {H.~C.}\ \bibnamefont {Lei}}, \bibinfo
  {author} {\bibfnamefont {J.~P.}\ \bibnamefont {Hu}}, \ and\ \bibinfo {author}
  {\bibfnamefont {J.-G.}\ \bibnamefont {Cheng}},\ }\href@noop {} {\bibfield
  {journal} {\bibinfo  {journal} {Phys. Rev. Lett.}\ }\textbf {\bibinfo
  {volume} {126}},\ \bibinfo {pages} {247001} (\bibinfo {year}
  {2021})}\BibitemShut {NoStop}%
\bibitem [{\citenamefont {Yu}\ \emph {et~al.}(2021)\citenamefont {Yu},
  \citenamefont {Ma}, \citenamefont {Zhuo}, \citenamefont {Liu}, \citenamefont
  {Wen}, \citenamefont {Lei}, \citenamefont {Ying},\ and\ \citenamefont
  {Chen}}]{Fu_2021}%
  \BibitemOpen
  \bibfield  {author} {\bibinfo {author} {\bibfnamefont {F.~H.}\ \bibnamefont
  {Yu}}, \bibinfo {author} {\bibfnamefont {D.~H.}\ \bibnamefont {Ma}}, \bibinfo
  {author} {\bibfnamefont {W.~Z.}\ \bibnamefont {Zhuo}}, \bibinfo {author}
  {\bibfnamefont {S.~Q.}\ \bibnamefont {Liu}}, \bibinfo {author} {\bibfnamefont
  {X.~K.}\ \bibnamefont {Wen}}, \bibinfo {author} {\bibfnamefont
  {B.}~\bibnamefont {Lei}}, \bibinfo {author} {\bibfnamefont {J.~J.}\
  \bibnamefont {Ying}}, \ and\ \bibinfo {author} {\bibfnamefont {X.~H.}\
  \bibnamefont {Chen}},\ }\href@noop {} {\bibfield  {journal} {\bibinfo
  {journal} {Nat. Commun.}\ }\textbf {\bibinfo {volume} {12}},\ \bibinfo
  {pages} {3645} (\bibinfo {year} {2021})}\BibitemShut {NoStop}%
\bibitem [{\citenamefont {Myers}\ \emph {et~al.}(1999)\citenamefont {Myers},
  \citenamefont {Bud'ko}, \citenamefont {Fisher}, \citenamefont {Islam},
  \citenamefont {Kleinke}, \citenamefont {Lacerda},\ and\ \citenamefont
  {Canfield}}]{Myers_1999a}%
  \BibitemOpen
  \bibfield  {author} {\bibinfo {author} {\bibfnamefont {K.~D.}\ \bibnamefont
  {Myers}}, \bibinfo {author} {\bibfnamefont {S.~L.}\ \bibnamefont {Bud'ko}},
  \bibinfo {author} {\bibfnamefont {I.~R.}\ \bibnamefont {Fisher}}, \bibinfo
  {author} {\bibfnamefont {Z.}~\bibnamefont {Islam}}, \bibinfo {author}
  {\bibfnamefont {H.}~\bibnamefont {Kleinke}}, \bibinfo {author} {\bibfnamefont
  {A.~H.}\ \bibnamefont {Lacerda}}, \ and\ \bibinfo {author} {\bibfnamefont
  {P.~C.}\ \bibnamefont {Canfield}},\ }\href@noop {} {\bibfield  {journal}
  {\bibinfo  {journal} {J. Magn. Magn. Mater.}\ }\textbf {\bibinfo {volume}
  {205}},\ \bibinfo {pages} {27} (\bibinfo {year} {1999})}\BibitemShut
  {NoStop}%
\bibitem [{\citenamefont {Song}\ \emph {et~al.}(2003)\citenamefont {Song},
  \citenamefont {Park}, \citenamefont {Koo}, \citenamefont {Lee}, \citenamefont
  {Rhee}, \citenamefont {Bud'ko}, \citenamefont {Canfield}, \citenamefont
  {Harmon},\ and\ \citenamefont {Goldman}}]{Song_2003}%
  \BibitemOpen
  \bibfield  {author} {\bibinfo {author} {\bibfnamefont {C.}~\bibnamefont
  {Song}}, \bibinfo {author} {\bibfnamefont {J.}~\bibnamefont {Park}}, \bibinfo
  {author} {\bibfnamefont {J.}~\bibnamefont {Koo}}, \bibinfo {author}
  {\bibfnamefont {K.-B.}\ \bibnamefont {Lee}}, \bibinfo {author} {\bibfnamefont
  {J.~Y.}\ \bibnamefont {Rhee}}, \bibinfo {author} {\bibfnamefont {S.~L.}\
  \bibnamefont {Bud'ko}}, \bibinfo {author} {\bibfnamefont {P.~C.}\
  \bibnamefont {Canfield}}, \bibinfo {author} {\bibfnamefont {B.~N.}\
  \bibnamefont {Harmon}}, \ and\ \bibinfo {author} {\bibfnamefont {A.~I.}\
  \bibnamefont {Goldman}},\ }\href@noop {} {\bibfield  {journal} {\bibinfo
  {journal} {Phys. Rev. B}\ }\textbf {\bibinfo {volume} {68}},\ \bibinfo
  {pages} {035113} (\bibinfo {year} {2003})}\BibitemShut {NoStop}%
\bibitem [{\citenamefont {Akiba}\ \emph {et~al.}(2021)\citenamefont {Akiba},
  \citenamefont {Nishimori}, \citenamefont {Umeshita},\ and\ \citenamefont
  {Kobayashi}}]{Akiba_2021}%
  \BibitemOpen
  \bibfield  {author} {\bibinfo {author} {\bibfnamefont {K.}~\bibnamefont
  {Akiba}}, \bibinfo {author} {\bibfnamefont {H.}~\bibnamefont {Nishimori}},
  \bibinfo {author} {\bibfnamefont {N.}~\bibnamefont {Umeshita}}, \ and\
  \bibinfo {author} {\bibfnamefont {T.~C.}\ \bibnamefont {Kobayashi}},\
  }\href@noop {} {\bibfield  {journal} {\bibinfo  {journal} {Phys. Rev. B}\
  }\textbf {\bibinfo {volume} {103}},\ \bibinfo {pages} {085134} (\bibinfo
  {year} {2021})}\BibitemShut {NoStop}%
\bibitem [{\citenamefont {Bosak}\ \emph {et~al.}(2021)\citenamefont {Bosak},
  \citenamefont {Souliou}, \citenamefont {Faugeras}, \citenamefont {Heid},
  \citenamefont {Molas}, \citenamefont {Chen}, \citenamefont {Wang},
  \citenamefont {Potemski},\ and\ \citenamefont {Le~Tacon}}]{Bosak_2021}%
  \BibitemOpen
  \bibfield  {author} {\bibinfo {author} {\bibfnamefont {A.}~\bibnamefont
  {Bosak}}, \bibinfo {author} {\bibfnamefont {S.-M.}\ \bibnamefont {Souliou}},
  \bibinfo {author} {\bibfnamefont {C.}~\bibnamefont {Faugeras}}, \bibinfo
  {author} {\bibfnamefont {R.}~\bibnamefont {Heid}}, \bibinfo {author}
  {\bibfnamefont {M.~R.}\ \bibnamefont {Molas}}, \bibinfo {author}
  {\bibfnamefont {R.-Y.}\ \bibnamefont {Chen}}, \bibinfo {author}
  {\bibfnamefont {N.-L.}\ \bibnamefont {Wang}}, \bibinfo {author}
  {\bibfnamefont {M.}~\bibnamefont {Potemski}}, \ and\ \bibinfo {author}
  {\bibfnamefont {M.}~\bibnamefont {Le~Tacon}},\ }\href@noop {} {\bibfield
  {journal} {\bibinfo  {journal} {Phys. Rev. Research}\ }\textbf {\bibinfo
  {volume} {3}},\ \bibinfo {pages} {033020} (\bibinfo {year}
  {2021})}\BibitemShut {NoStop}%
\bibitem [{\citenamefont {Akiba}\ \emph {et~al.}(2022)\citenamefont {Akiba},
  \citenamefont {Umeshita},\ and\ \citenamefont {Kobayashi}}]{Akiba_2022}%
  \BibitemOpen
  \bibfield  {author} {\bibinfo {author} {\bibfnamefont {K.}~\bibnamefont
  {Akiba}}, \bibinfo {author} {\bibfnamefont {N.}~\bibnamefont {Umeshita}}, \
  and\ \bibinfo {author} {\bibfnamefont {T.~C.}\ \bibnamefont {Kobayashi}},\
  }\href@noop {} {\bibfield  {journal} {\bibinfo  {journal} {Phys. Rev. B}\
  }\textbf {\bibinfo {volume} {105}},\ \bibinfo {pages} {035108} (\bibinfo
  {year} {2022})}\BibitemShut {NoStop}%
\bibitem [{\citenamefont {Wang}\ and\ \citenamefont
  {Petrovic}(2012)}]{Wang_2012Oct}%
  \BibitemOpen
  \bibfield  {author} {\bibinfo {author} {\bibfnamefont {K.}~\bibnamefont
  {Wang}}\ and\ \bibinfo {author} {\bibfnamefont {C.}~\bibnamefont
  {Petrovic}},\ }\href@noop {} {\bibfield  {journal} {\bibinfo  {journal}
  {Phys. Rev. B}\ }\textbf {\bibinfo {volume} {86}},\ \bibinfo {pages} {155213}
  (\bibinfo {year} {2012})}\BibitemShut {NoStop}%
\bibitem [{\citenamefont {Shi}\ \emph {et~al.}(2016)\citenamefont {Shi},
  \citenamefont {Richard}, \citenamefont {Wang}, \citenamefont {Liu},
  \citenamefont {Matt}, \citenamefont {Xu}, \citenamefont {Dhaka},
  \citenamefont {Ristic}, \citenamefont {Qian}, \citenamefont {Yang},
  \citenamefont {Petrovic}, \citenamefont {Shi},\ and\ \citenamefont
  {Ding}}]{Shi_2016}%
  \BibitemOpen
  \bibfield  {author} {\bibinfo {author} {\bibfnamefont {X.}~\bibnamefont
  {Shi}}, \bibinfo {author} {\bibfnamefont {P.}~\bibnamefont {Richard}},
  \bibinfo {author} {\bibfnamefont {K.}~\bibnamefont {Wang}}, \bibinfo {author}
  {\bibfnamefont {M.}~\bibnamefont {Liu}}, \bibinfo {author} {\bibfnamefont
  {C.~E.}\ \bibnamefont {Matt}}, \bibinfo {author} {\bibfnamefont
  {N.}~\bibnamefont {Xu}}, \bibinfo {author} {\bibfnamefont {R.~S.}\
  \bibnamefont {Dhaka}}, \bibinfo {author} {\bibfnamefont {Z.}~\bibnamefont
  {Ristic}}, \bibinfo {author} {\bibfnamefont {T.}~\bibnamefont {Qian}},
  \bibinfo {author} {\bibfnamefont {Y.-F.}\ \bibnamefont {Yang}}, \bibinfo
  {author} {\bibfnamefont {C.}~\bibnamefont {Petrovic}}, \bibinfo {author}
  {\bibfnamefont {M.}~\bibnamefont {Shi}}, \ and\ \bibinfo {author}
  {\bibfnamefont {H.}~\bibnamefont {Ding}},\ }\href@noop {} {\bibfield
  {journal} {\bibinfo  {journal} {Phys. Rev. B}\ }\textbf {\bibinfo {volume}
  {93}},\ \bibinfo {pages} {081105(R)} (\bibinfo {year} {2016})}\BibitemShut
  {NoStop}%
\bibitem [{\citenamefont {Hoffmann}(1987)}]{Hoffmann_1987}%
  \BibitemOpen
  \bibfield  {author} {\bibinfo {author} {\bibfnamefont {R.}~\bibnamefont
  {Hoffmann}},\ }\href@noop {} {\bibfield  {journal} {\bibinfo  {journal}
  {Angew. Chem. Int. Ed. Engl.}\ }\textbf {\bibinfo {volume} {26}},\ \bibinfo
  {pages} {846} (\bibinfo {year} {1987})}\BibitemShut {NoStop}%
\bibitem [{\citenamefont {Klemenz}\ \emph {et~al.}(2019)\citenamefont
  {Klemenz}, \citenamefont {Lei},\ and\ \citenamefont {Schoop}}]{Klemenz_2019}%
  \BibitemOpen
  \bibfield  {author} {\bibinfo {author} {\bibfnamefont {S.}~\bibnamefont
  {Klemenz}}, \bibinfo {author} {\bibfnamefont {S.}~\bibnamefont {Lei}}, \ and\
  \bibinfo {author} {\bibfnamefont {L.~M.}\ \bibnamefont {Schoop}},\
  }\href@noop {} {\bibfield  {journal} {\bibinfo  {journal} {Annu. Rev. Mater.
  Res.}\ }\textbf {\bibinfo {volume} {49}},\ \bibinfo {pages} {185} (\bibinfo
  {year} {2019})}\BibitemShut {NoStop}%
\bibitem [{\citenamefont {Klemenz}\ \emph {et~al.}(2020)\citenamefont
  {Klemenz}, \citenamefont {Hay}, \citenamefont {Teicher}, \citenamefont
  {Topp}, \citenamefont {Cano},\ and\ \citenamefont {Schoop}}]{Klemenz_2020}%
  \BibitemOpen
  \bibfield  {author} {\bibinfo {author} {\bibfnamefont {S.}~\bibnamefont
  {Klemenz}}, \bibinfo {author} {\bibfnamefont {A.~K.}\ \bibnamefont {Hay}},
  \bibinfo {author} {\bibfnamefont {S.~M.~L.}\ \bibnamefont {Teicher}},
  \bibinfo {author} {\bibfnamefont {A.}~\bibnamefont {Topp}}, \bibinfo {author}
  {\bibfnamefont {J.}~\bibnamefont {Cano}}, \ and\ \bibinfo {author}
  {\bibfnamefont {L.~M.}\ \bibnamefont {Schoop}},\ }\href@noop {} {\bibfield
  {journal} {\bibinfo  {journal} {J. Am. Chem. Soc.}\ }\textbf {\bibinfo
  {volume} {142}},\ \bibinfo {pages} {6350} (\bibinfo {year}
  {2020})}\BibitemShut {NoStop}%
\bibitem [{\citenamefont {Du}\ \emph {et~al.}(2020)\citenamefont {Du},
  \citenamefont {Su}, \citenamefont {Luo}, \citenamefont {Shen}, \citenamefont
  {Nie}, \citenamefont {Yin}, \citenamefont {Chen}, \citenamefont {Li},
  \citenamefont {Smidman},\ and\ \citenamefont {Yuan}}]{Du_2020}%
  \BibitemOpen
  \bibfield  {author} {\bibinfo {author} {\bibfnamefont {F.}~\bibnamefont
  {Du}}, \bibinfo {author} {\bibfnamefont {H.}~\bibnamefont {Su}}, \bibinfo
  {author} {\bibfnamefont {S.~S.}\ \bibnamefont {Luo}}, \bibinfo {author}
  {\bibfnamefont {B.}~\bibnamefont {Shen}}, \bibinfo {author} {\bibfnamefont
  {Z.~Y.}\ \bibnamefont {Nie}}, \bibinfo {author} {\bibfnamefont {L.~C.}\
  \bibnamefont {Yin}}, \bibinfo {author} {\bibfnamefont {Y.}~\bibnamefont
  {Chen}}, \bibinfo {author} {\bibfnamefont {R.}~\bibnamefont {Li}}, \bibinfo
  {author} {\bibfnamefont {M.}~\bibnamefont {Smidman}}, \ and\ \bibinfo
  {author} {\bibfnamefont {H.~Q.}\ \bibnamefont {Yuan}},\ }\href@noop {}
  {\bibfield  {journal} {\bibinfo  {journal} {Phys. Rev. B}\ }\textbf {\bibinfo
  {volume} {102}},\ \bibinfo {pages} {144510} (\bibinfo {year}
  {2020})}\BibitemShut {NoStop}%
\bibitem [{\citenamefont {Muro}\ \emph {et~al.}(1997)\citenamefont {Muro},
  \citenamefont {Takeda},\ and\ \citenamefont {Ishikawa}}]{Muro_1997}%
  \BibitemOpen
  \bibfield  {author} {\bibinfo {author} {\bibfnamefont {Y.}~\bibnamefont
  {Muro}}, \bibinfo {author} {\bibfnamefont {N.}~\bibnamefont {Takeda}}, \ and\
  \bibinfo {author} {\bibfnamefont {M.}~\bibnamefont {Ishikawa}},\ }\href@noop
  {} {\bibfield  {journal} {\bibinfo  {journal} {J. Alloy. Compd.}\ }\textbf
  {\bibinfo {volume} {257}},\ \bibinfo {pages} {23} (\bibinfo {year}
  {1997})}\BibitemShut {NoStop}%
\bibitem [{\citenamefont {Kobayashi}\ \emph {et~al.}(2007)\citenamefont
  {Kobayashi}, \citenamefont {Hidaka}, \citenamefont {Kotegawa}, \citenamefont
  {Fujiwara},\ and\ \citenamefont {Eremets}}]{Kobayashi_2007}%
  \BibitemOpen
  \bibfield  {author} {\bibinfo {author} {\bibfnamefont {T.~C.}\ \bibnamefont
  {Kobayashi}}, \bibinfo {author} {\bibfnamefont {H.}~\bibnamefont {Hidaka}},
  \bibinfo {author} {\bibfnamefont {H.}~\bibnamefont {Kotegawa}}, \bibinfo
  {author} {\bibfnamefont {K.}~\bibnamefont {Fujiwara}}, \ and\ \bibinfo
  {author} {\bibfnamefont {M.~I.}\ \bibnamefont {Eremets}},\ }\href@noop {}
  {\bibfield  {journal} {\bibinfo  {journal} {Rev. Sci. Instrum}\ }\textbf
  {\bibinfo {volume} {78}},\ \bibinfo {pages} {023909} (\bibinfo {year}
  {2007})}\BibitemShut {NoStop}%
\bibitem [{\citenamefont {Kitagawa}\ \emph {et~al.}(2010)\citenamefont
  {Kitagawa}, \citenamefont {Gotou}, \citenamefont {Yagi}, \citenamefont
  {Yamada}, \citenamefont {Matsumoto}, \citenamefont {Uwatoko},\ and\
  \citenamefont {Takigawa}}]{Kitagawa_2010}%
  \BibitemOpen
  \bibfield  {author} {\bibinfo {author} {\bibfnamefont {K.}~\bibnamefont
  {Kitagawa}}, \bibinfo {author} {\bibfnamefont {H.}~\bibnamefont {Gotou}},
  \bibinfo {author} {\bibfnamefont {T.}~\bibnamefont {Yagi}}, \bibinfo {author}
  {\bibfnamefont {A.}~\bibnamefont {Yamada}}, \bibinfo {author} {\bibfnamefont
  {T.}~\bibnamefont {Matsumoto}}, \bibinfo {author} {\bibfnamefont
  {Y.}~\bibnamefont {Uwatoko}}, \ and\ \bibinfo {author} {\bibfnamefont
  {M.}~\bibnamefont {Takigawa}},\ }\href@noop {} {\bibfield  {journal}
  {\bibinfo  {journal} {J. Phys. Soc. Jpn.}\ }\textbf {\bibinfo {volume}
  {79}},\ \bibinfo {pages} {024001} (\bibinfo {year} {2010})}\BibitemShut
  {NoStop}%
\bibitem [{\citenamefont {Murata}\ \emph {et~al.}(2008)\citenamefont {Murata},
  \citenamefont {Yokogawa}, \citenamefont {Yoshino}, \citenamefont {Klotz},
  \citenamefont {Munsch}, \citenamefont {Irizawa}, \citenamefont {Nishiyama},
  \citenamefont {Iizuka}, \citenamefont {Nanba}, \citenamefont {Okada},
  \citenamefont {Shiraga},\ and\ \citenamefont {Aoyama}}]{Murata_2008}%
  \BibitemOpen
  \bibfield  {author} {\bibinfo {author} {\bibfnamefont {K.}~\bibnamefont
  {Murata}}, \bibinfo {author} {\bibfnamefont {K.}~\bibnamefont {Yokogawa}},
  \bibinfo {author} {\bibfnamefont {H.}~\bibnamefont {Yoshino}}, \bibinfo
  {author} {\bibfnamefont {S.}~\bibnamefont {Klotz}}, \bibinfo {author}
  {\bibfnamefont {P.}~\bibnamefont {Munsch}}, \bibinfo {author} {\bibfnamefont
  {A.}~\bibnamefont {Irizawa}}, \bibinfo {author} {\bibfnamefont
  {M.}~\bibnamefont {Nishiyama}}, \bibinfo {author} {\bibfnamefont
  {K.}~\bibnamefont {Iizuka}}, \bibinfo {author} {\bibfnamefont
  {T.}~\bibnamefont {Nanba}}, \bibinfo {author} {\bibfnamefont
  {T.}~\bibnamefont {Okada}}, \bibinfo {author} {\bibfnamefont
  {Y.}~\bibnamefont {Shiraga}}, \ and\ \bibinfo {author} {\bibfnamefont
  {S.}~\bibnamefont {Aoyama}},\ }\href@noop {} {\bibfield  {journal} {\bibinfo
  {journal} {Rev. Sci. Instrum.}\ }\textbf {\bibinfo {volume} {79}},\ \bibinfo
  {pages} {085101} (\bibinfo {year} {2008})}\BibitemShut {NoStop}%
\bibitem [{SM_()}]{SM_URL}%
  \BibitemOpen
  \href@noop {} {}\bibinfo {note} {See Supplemental Material at
  [URL]}\BibitemShut {NoStop}%
\bibitem [{\citenamefont {Giannozzi}\ \emph {et~al.}(2009)\citenamefont
  {Giannozzi}, \citenamefont {Baroni}, \citenamefont {Bonini}, \citenamefont
  {Calandra}, \citenamefont {Car}, \citenamefont {Cavazzoni}, \citenamefont
  {Ceresoli}, \citenamefont {Chiarotti}, \citenamefont {Cococcioni},
  \citenamefont {Dabo}, \citenamefont {Corso}, \citenamefont {de~Gironcoli},
  \citenamefont {Fabris}, \citenamefont {Fratesi}, \citenamefont {Gebauer},
  \citenamefont {Gerstmann}, \citenamefont {Gougoussis}, \citenamefont
  {Kokalj}, \citenamefont {Lazzeri}, \citenamefont {Martin-Samos},
  \citenamefont {Marzari}, \citenamefont {Mauri}, \citenamefont {Mazzarello},
  \citenamefont {Paolini}, \citenamefont {Pasquarello}, \citenamefont
  {Paulatto}, \citenamefont {Sbraccia}, \citenamefont {Scandolo}, \citenamefont
  {Sclauzero}, \citenamefont {Seitsonen}, \citenamefont {Smogunov},
  \citenamefont {Umari},\ and\ \citenamefont {Wentzcovitch}}]{Giannozzi_2009}%
  \BibitemOpen
  \bibfield  {author} {\bibinfo {author} {\bibfnamefont {P.}~\bibnamefont
  {Giannozzi}}, \bibinfo {author} {\bibfnamefont {S.}~\bibnamefont {Baroni}},
  \bibinfo {author} {\bibfnamefont {N.}~\bibnamefont {Bonini}}, \bibinfo
  {author} {\bibfnamefont {M.}~\bibnamefont {Calandra}}, \bibinfo {author}
  {\bibfnamefont {R.}~\bibnamefont {Car}}, \bibinfo {author} {\bibfnamefont
  {C.}~\bibnamefont {Cavazzoni}}, \bibinfo {author} {\bibfnamefont
  {D.}~\bibnamefont {Ceresoli}}, \bibinfo {author} {\bibfnamefont {G.~L.}\
  \bibnamefont {Chiarotti}}, \bibinfo {author} {\bibfnamefont {M.}~\bibnamefont
  {Cococcioni}}, \bibinfo {author} {\bibfnamefont {I.}~\bibnamefont {Dabo}},
  \bibinfo {author} {\bibfnamefont {A.~D.}\ \bibnamefont {Corso}}, \bibinfo
  {author} {\bibfnamefont {S.}~\bibnamefont {de~Gironcoli}}, \bibinfo {author}
  {\bibfnamefont {S.}~\bibnamefont {Fabris}}, \bibinfo {author} {\bibfnamefont
  {G.}~\bibnamefont {Fratesi}}, \bibinfo {author} {\bibfnamefont
  {R.}~\bibnamefont {Gebauer}}, \bibinfo {author} {\bibfnamefont
  {U.}~\bibnamefont {Gerstmann}}, \bibinfo {author} {\bibfnamefont
  {C.}~\bibnamefont {Gougoussis}}, \bibinfo {author} {\bibfnamefont
  {A.}~\bibnamefont {Kokalj}}, \bibinfo {author} {\bibfnamefont
  {M.}~\bibnamefont {Lazzeri}}, \bibinfo {author} {\bibfnamefont
  {L.}~\bibnamefont {Martin-Samos}}, \bibinfo {author} {\bibfnamefont
  {N.}~\bibnamefont {Marzari}}, \bibinfo {author} {\bibfnamefont
  {F.}~\bibnamefont {Mauri}}, \bibinfo {author} {\bibfnamefont
  {R.}~\bibnamefont {Mazzarello}}, \bibinfo {author} {\bibfnamefont
  {S.}~\bibnamefont {Paolini}}, \bibinfo {author} {\bibfnamefont
  {A.}~\bibnamefont {Pasquarello}}, \bibinfo {author} {\bibfnamefont
  {L.}~\bibnamefont {Paulatto}}, \bibinfo {author} {\bibfnamefont
  {C.}~\bibnamefont {Sbraccia}}, \bibinfo {author} {\bibfnamefont
  {S.}~\bibnamefont {Scandolo}}, \bibinfo {author} {\bibfnamefont
  {G.}~\bibnamefont {Sclauzero}}, \bibinfo {author} {\bibfnamefont {A.~P.}\
  \bibnamefont {Seitsonen}}, \bibinfo {author} {\bibfnamefont {A.}~\bibnamefont
  {Smogunov}}, \bibinfo {author} {\bibfnamefont {P.}~\bibnamefont {Umari}}, \
  and\ \bibinfo {author} {\bibfnamefont {R.~M.}\ \bibnamefont {Wentzcovitch}},\
  }\href@noop {} {\bibfield  {journal} {\bibinfo  {journal} {J. Phys.: Condens.
  Matter}\ }\textbf {\bibinfo {volume} {21}},\ \bibinfo {pages} {395502}
  (\bibinfo {year} {2009})}\BibitemShut {NoStop}%
\bibitem [{\citenamefont {Giannozzi}\ \emph {et~al.}(2017)\citenamefont
  {Giannozzi}, \citenamefont {Andreussi}, \citenamefont {Brumme}, \citenamefont
  {Bunau}, \citenamefont {{Buongiorno Nardelli}}, \citenamefont {Calandra},
  \citenamefont {Car}, \citenamefont {Cavazzoni}, \citenamefont {Ceresoli},
  \citenamefont {Cococcioni}, \citenamefont {Colonna}, \citenamefont
  {Carnimeo}, \citenamefont {{Dal Corso}}, \citenamefont {{De Gironcoli}},
  \citenamefont {Delugas}, \citenamefont {Distasio}, \citenamefont {Ferretti},
  \citenamefont {Floris}, \citenamefont {Fratesi}, \citenamefont {Fugallo},
  \citenamefont {Gebauer}, \citenamefont {Gerstmann}, \citenamefont {Giustino},
  \citenamefont {Gorni}, \citenamefont {Jia}, \citenamefont {Kawamura},
  \citenamefont {Ko}, \citenamefont {Kokalj}, \citenamefont
  {K{\"u}c{\"u}kbenli}, \citenamefont {Lazzeri}, \citenamefont {Marsili},
  \citenamefont {Marzari}, \citenamefont {Mauri}, \citenamefont {Nguyen},
  \citenamefont {Nguyen}, \citenamefont {Otero-De-La-Roza}, \citenamefont
  {Paulatto}, \citenamefont {Ponc{\'e}}, \citenamefont {Rocca}, \citenamefont
  {Sabatini}, \citenamefont {Santra}, \citenamefont {Schlipf}, \citenamefont
  {Seitsonen}, \citenamefont {Smogunov}, \citenamefont {Timrov}, \citenamefont
  {Thonhauser}, \citenamefont {Umari}, \citenamefont {Vast}, \citenamefont
  {Wu},\ and\ \citenamefont {Baroni}}]{Giannozzi_2017}%
  \BibitemOpen
  \bibfield  {author} {\bibinfo {author} {\bibfnamefont {P.}~\bibnamefont
  {Giannozzi}}, \bibinfo {author} {\bibfnamefont {O.}~\bibnamefont
  {Andreussi}}, \bibinfo {author} {\bibfnamefont {T.}~\bibnamefont {Brumme}},
  \bibinfo {author} {\bibfnamefont {O.}~\bibnamefont {Bunau}}, \bibinfo
  {author} {\bibfnamefont {M.}~\bibnamefont {{Buongiorno Nardelli}}}, \bibinfo
  {author} {\bibfnamefont {M.}~\bibnamefont {Calandra}}, \bibinfo {author}
  {\bibfnamefont {R.}~\bibnamefont {Car}}, \bibinfo {author} {\bibfnamefont
  {C.}~\bibnamefont {Cavazzoni}}, \bibinfo {author} {\bibfnamefont
  {D.}~\bibnamefont {Ceresoli}}, \bibinfo {author} {\bibfnamefont
  {M.}~\bibnamefont {Cococcioni}}, \bibinfo {author} {\bibfnamefont
  {N.}~\bibnamefont {Colonna}}, \bibinfo {author} {\bibfnamefont
  {I.}~\bibnamefont {Carnimeo}}, \bibinfo {author} {\bibfnamefont
  {A.}~\bibnamefont {{Dal Corso}}}, \bibinfo {author} {\bibfnamefont
  {S.}~\bibnamefont {{De Gironcoli}}}, \bibinfo {author} {\bibfnamefont
  {P.}~\bibnamefont {Delugas}}, \bibinfo {author} {\bibfnamefont
  {R.}~\bibnamefont {Distasio}}, \bibinfo {author} {\bibfnamefont
  {A.}~\bibnamefont {Ferretti}}, \bibinfo {author} {\bibfnamefont
  {A.}~\bibnamefont {Floris}}, \bibinfo {author} {\bibfnamefont
  {G.}~\bibnamefont {Fratesi}}, \bibinfo {author} {\bibfnamefont
  {G.}~\bibnamefont {Fugallo}}, \bibinfo {author} {\bibfnamefont
  {R.}~\bibnamefont {Gebauer}}, \bibinfo {author} {\bibfnamefont
  {U.}~\bibnamefont {Gerstmann}}, \bibinfo {author} {\bibfnamefont
  {F.}~\bibnamefont {Giustino}}, \bibinfo {author} {\bibfnamefont
  {T.}~\bibnamefont {Gorni}}, \bibinfo {author} {\bibfnamefont
  {J.}~\bibnamefont {Jia}}, \bibinfo {author} {\bibfnamefont {M.}~\bibnamefont
  {Kawamura}}, \bibinfo {author} {\bibfnamefont {H.}~\bibnamefont {Ko}},
  \bibinfo {author} {\bibfnamefont {A.}~\bibnamefont {Kokalj}}, \bibinfo
  {author} {\bibfnamefont {E.}~\bibnamefont {K{\"u}c{\"u}kbenli}}, \bibinfo
  {author} {\bibfnamefont {M.}~\bibnamefont {Lazzeri}}, \bibinfo {author}
  {\bibfnamefont {M.}~\bibnamefont {Marsili}}, \bibinfo {author} {\bibfnamefont
  {N.}~\bibnamefont {Marzari}}, \bibinfo {author} {\bibfnamefont
  {F.}~\bibnamefont {Mauri}}, \bibinfo {author} {\bibfnamefont
  {N.}~\bibnamefont {Nguyen}}, \bibinfo {author} {\bibfnamefont
  {H.}~\bibnamefont {Nguyen}}, \bibinfo {author} {\bibfnamefont
  {A.}~\bibnamefont {Otero-De-La-Roza}}, \bibinfo {author} {\bibfnamefont
  {L.}~\bibnamefont {Paulatto}}, \bibinfo {author} {\bibfnamefont
  {S.}~\bibnamefont {Ponc{\'e}}}, \bibinfo {author} {\bibfnamefont
  {D.}~\bibnamefont {Rocca}}, \bibinfo {author} {\bibfnamefont
  {R.}~\bibnamefont {Sabatini}}, \bibinfo {author} {\bibfnamefont
  {B.}~\bibnamefont {Santra}}, \bibinfo {author} {\bibfnamefont
  {M.}~\bibnamefont {Schlipf}}, \bibinfo {author} {\bibfnamefont
  {A.}~\bibnamefont {Seitsonen}}, \bibinfo {author} {\bibfnamefont
  {A.}~\bibnamefont {Smogunov}}, \bibinfo {author} {\bibfnamefont
  {I.}~\bibnamefont {Timrov}}, \bibinfo {author} {\bibfnamefont
  {T.}~\bibnamefont {Thonhauser}}, \bibinfo {author} {\bibfnamefont
  {P.}~\bibnamefont {Umari}}, \bibinfo {author} {\bibfnamefont
  {N.}~\bibnamefont {Vast}}, \bibinfo {author} {\bibfnamefont {X.}~\bibnamefont
  {Wu}}, \ and\ \bibinfo {author} {\bibfnamefont {S.}~\bibnamefont {Baroni}},\
  }\href@noop {} {\bibfield  {journal} {\bibinfo  {journal} {J. Phys.: Condens.
  Matter}\ }\textbf {\bibinfo {volume} {29}},\ \bibinfo {pages} {465901}
  (\bibinfo {year} {2017})}\BibitemShut {NoStop}%
\bibitem [{\citenamefont {Kawamura}\ \emph {et~al.}(2014)\citenamefont
  {Kawamura}, \citenamefont {Gohda},\ and\ \citenamefont
  {Tsuneyuki}}]{Kawamura_2014}%
  \BibitemOpen
  \bibfield  {author} {\bibinfo {author} {\bibfnamefont {M.}~\bibnamefont
  {Kawamura}}, \bibinfo {author} {\bibfnamefont {Y.}~\bibnamefont {Gohda}}, \
  and\ \bibinfo {author} {\bibfnamefont {S.}~\bibnamefont {Tsuneyuki}},\
  }\href@noop {} {\bibfield  {journal} {\bibinfo  {journal} {Phys. Rev. B}\
  }\textbf {\bibinfo {volume} {89}},\ \bibinfo {pages} {094515} (\bibinfo
  {year} {2014})}\BibitemShut {NoStop}%
\bibitem [{\citenamefont {Pizzi}\ \emph {et~al.}(2020)\citenamefont {Pizzi},
  \citenamefont {Vitale}, \citenamefont {Arita}, \citenamefont {Bluegel},
  \citenamefont {Freimuth}, \citenamefont {G{\'e}ranton}, \citenamefont
  {Gibertini}, \citenamefont {Gresch}, \citenamefont {Johnson}, \citenamefont
  {Koretsune}, \citenamefont {Ibanez}, \citenamefont {Lee}, \citenamefont
  {Lihm}, \citenamefont {Marchand}, \citenamefont {Marrazzo}, \citenamefont
  {Mokrousov}, \citenamefont {Mustafa}, \citenamefont {Nohara}, \citenamefont
  {Nomura}, \citenamefont {Paulatto}, \citenamefont {Ponc{\'e}}, \citenamefont
  {Ponweiser}, \citenamefont {Qiao}, \citenamefont {Th{\"o}le}, \citenamefont
  {Tsirkin}, \citenamefont {Wierzbowska}, \citenamefont {Marzari},
  \citenamefont {Vanderbilt}, \citenamefont {Souza}, \citenamefont {Mostofi},\
  and\ \citenamefont {Yates}}]{Pizzi_2020}%
  \BibitemOpen
  \bibfield  {author} {\bibinfo {author} {\bibfnamefont {G.}~\bibnamefont
  {Pizzi}}, \bibinfo {author} {\bibfnamefont {V.}~\bibnamefont {Vitale}},
  \bibinfo {author} {\bibfnamefont {R.}~\bibnamefont {Arita}}, \bibinfo
  {author} {\bibfnamefont {S.}~\bibnamefont {Bluegel}}, \bibinfo {author}
  {\bibfnamefont {F.}~\bibnamefont {Freimuth}}, \bibinfo {author}
  {\bibfnamefont {G.}~\bibnamefont {G{\'e}ranton}}, \bibinfo {author}
  {\bibfnamefont {M.}~\bibnamefont {Gibertini}}, \bibinfo {author}
  {\bibfnamefont {D.}~\bibnamefont {Gresch}}, \bibinfo {author} {\bibfnamefont
  {C.}~\bibnamefont {Johnson}}, \bibinfo {author} {\bibfnamefont
  {T.}~\bibnamefont {Koretsune}}, \bibinfo {author} {\bibfnamefont
  {J.}~\bibnamefont {Ibanez}}, \bibinfo {author} {\bibfnamefont
  {H.}~\bibnamefont {Lee}}, \bibinfo {author} {\bibfnamefont {J.-M.}\
  \bibnamefont {Lihm}}, \bibinfo {author} {\bibfnamefont {D.}~\bibnamefont
  {Marchand}}, \bibinfo {author} {\bibfnamefont {A.}~\bibnamefont {Marrazzo}},
  \bibinfo {author} {\bibfnamefont {Y.}~\bibnamefont {Mokrousov}}, \bibinfo
  {author} {\bibfnamefont {J.~I.}\ \bibnamefont {Mustafa}}, \bibinfo {author}
  {\bibfnamefont {Y.}~\bibnamefont {Nohara}}, \bibinfo {author} {\bibfnamefont
  {Y.}~\bibnamefont {Nomura}}, \bibinfo {author} {\bibfnamefont
  {L.}~\bibnamefont {Paulatto}}, \bibinfo {author} {\bibfnamefont
  {S.}~\bibnamefont {Ponc{\'e}}}, \bibinfo {author} {\bibfnamefont
  {T.}~\bibnamefont {Ponweiser}}, \bibinfo {author} {\bibfnamefont
  {J.}~\bibnamefont {Qiao}}, \bibinfo {author} {\bibfnamefont {F.}~\bibnamefont
  {Th{\"o}le}}, \bibinfo {author} {\bibfnamefont {S.~S.}\ \bibnamefont
  {Tsirkin}}, \bibinfo {author} {\bibfnamefont {M.}~\bibnamefont
  {Wierzbowska}}, \bibinfo {author} {\bibfnamefont {N.}~\bibnamefont
  {Marzari}}, \bibinfo {author} {\bibfnamefont {D.}~\bibnamefont {Vanderbilt}},
  \bibinfo {author} {\bibfnamefont {I.}~\bibnamefont {Souza}}, \bibinfo
  {author} {\bibfnamefont {A.~A.}\ \bibnamefont {Mostofi}}, \ and\ \bibinfo
  {author} {\bibfnamefont {J.~R.}\ \bibnamefont {Yates}},\ }\href@noop {}
  {\bibfield  {journal} {\bibinfo  {journal} {J. Phys.: Condens. Matter}\
  }\textbf {\bibinfo {volume} {32}},\ \bibinfo {pages} {165902} (\bibinfo
  {year} {2020})}\BibitemShut {NoStop}%
\bibitem [{\citenamefont {Ponc\'e}\ \emph {et~al.}(2016)\citenamefont
  {Ponc\'e}, \citenamefont {Margine}, \citenamefont {Verdi},\ and\
  \citenamefont {Giustino}}]{Ponce_2016}%
  \BibitemOpen
  \bibfield  {author} {\bibinfo {author} {\bibfnamefont {S.}~\bibnamefont
  {Ponc\'e}}, \bibinfo {author} {\bibfnamefont {E.}~\bibnamefont {Margine}},
  \bibinfo {author} {\bibfnamefont {C.}~\bibnamefont {Verdi}}, \ and\ \bibinfo
  {author} {\bibfnamefont {F.}~\bibnamefont {Giustino}},\ }\href@noop {}
  {\bibfield  {journal} {\bibinfo  {journal} {Comput. Phys. Commun.}\ }\textbf
  {\bibinfo {volume} {209}},\ \bibinfo {pages} {116} (\bibinfo {year}
  {2016})}\BibitemShut {NoStop}%
\bibitem [{\citenamefont {Kawamura}(2019)}]{Kawamura_2019}%
  \BibitemOpen
  \bibfield  {author} {\bibinfo {author} {\bibfnamefont {M.}~\bibnamefont
  {Kawamura}},\ }\href@noop {} {\bibfield  {journal} {\bibinfo  {journal}
  {Comput. Phys. Commun.}\ }\textbf {\bibinfo {volume} {239}},\ \bibinfo
  {pages} {197} (\bibinfo {year} {2019})}\BibitemShut {NoStop}%
\bibitem [{\citenamefont {Inada}\ \emph {et~al.}(2002)\citenamefont {Inada},
  \citenamefont {Thamizhavel}, \citenamefont {Yamagami}, \citenamefont
  {Takeuchi}, \citenamefont {Sawai}, \citenamefont {Ikeda}, \citenamefont
  {Shishido}, \citenamefont {Okubo}, \citenamefont {Yamada}, \citenamefont
  {Sugiyama}, \citenamefont {Nakamura}, \citenamefont {Yamamoto}, \citenamefont
  {Kindo}, \citenamefont {Ebihara}, \citenamefont {Galatanu}, \citenamefont
  {Yamamoto}, \citenamefont {Settai},\ and\ \citenamefont
  {Onuki}}]{Inada_2002}%
  \BibitemOpen
  \bibfield  {author} {\bibinfo {author} {\bibfnamefont {Y.}~\bibnamefont
  {Inada}}, \bibinfo {author} {\bibfnamefont {A.}~\bibnamefont {Thamizhavel}},
  \bibinfo {author} {\bibfnamefont {H.}~\bibnamefont {Yamagami}}, \bibinfo
  {author} {\bibfnamefont {T.}~\bibnamefont {Takeuchi}}, \bibinfo {author}
  {\bibfnamefont {Y.}~\bibnamefont {Sawai}}, \bibinfo {author} {\bibfnamefont
  {S.}~\bibnamefont {Ikeda}}, \bibinfo {author} {\bibfnamefont
  {H.}~\bibnamefont {Shishido}}, \bibinfo {author} {\bibfnamefont
  {T.}~\bibnamefont {Okubo}}, \bibinfo {author} {\bibfnamefont
  {M.}~\bibnamefont {Yamada}}, \bibinfo {author} {\bibfnamefont
  {K.}~\bibnamefont {Sugiyama}}, \bibinfo {author} {\bibfnamefont
  {N.}~\bibnamefont {Nakamura}}, \bibinfo {author} {\bibfnamefont
  {T.}~\bibnamefont {Yamamoto}}, \bibinfo {author} {\bibfnamefont
  {K.}~\bibnamefont {Kindo}}, \bibinfo {author} {\bibfnamefont
  {T.}~\bibnamefont {Ebihara}}, \bibinfo {author} {\bibfnamefont
  {A.}~\bibnamefont {Galatanu}}, \bibinfo {author} {\bibfnamefont
  {E.}~\bibnamefont {Yamamoto}}, \bibinfo {author} {\bibfnamefont
  {R.}~\bibnamefont {Settai}}, \ and\ \bibinfo {author} {\bibfnamefont
  {Y.}~\bibnamefont {Onuki}},\ }\href@noop {} {\bibfield  {journal} {\bibinfo
  {journal} {Phil. Mag.}\ }\textbf {\bibinfo {volume} {82}},\ \bibinfo {pages}
  {1867} (\bibinfo {year} {2002})}\BibitemShut {NoStop}%
\bibitem [{\citenamefont {McMillan}(1968)}]{McMillan_1968}%
  \BibitemOpen
  \bibfield  {author} {\bibinfo {author} {\bibfnamefont {W.~L.}\ \bibnamefont
  {McMillan}},\ }\href@noop {} {\bibfield  {journal} {\bibinfo  {journal}
  {Phys. Rev.}\ }\textbf {\bibinfo {volume} {167}},\ \bibinfo {pages} {331}
  (\bibinfo {year} {1968})}\BibitemShut {NoStop}%
\bibitem [{\citenamefont {Dynes}(1972)}]{Dynes_1972}%
  \BibitemOpen
  \bibfield  {author} {\bibinfo {author} {\bibfnamefont {R.~C.}\ \bibnamefont
  {Dynes}},\ }\href@noop {} {\bibfield  {journal} {\bibinfo  {journal} {Solid
  State Commun.}\ }\textbf {\bibinfo {volume} {10}},\ \bibinfo {pages} {615}
  (\bibinfo {year} {1972})}\BibitemShut {NoStop}%
\bibitem [{\citenamefont {Allen}\ and\ \citenamefont
  {Dynes}(1975)}]{Allen_1975}%
  \BibitemOpen
  \bibfield  {author} {\bibinfo {author} {\bibfnamefont {P.~B.}\ \bibnamefont
  {Allen}}\ and\ \bibinfo {author} {\bibfnamefont {R.~C.}\ \bibnamefont
  {Dynes}},\ }\href@noop {} {\bibfield  {journal} {\bibinfo  {journal} {Phys.
  Rev. B}\ }\textbf {\bibinfo {volume} {12}},\ \bibinfo {pages} {905} (\bibinfo
  {year} {1975})}\BibitemShut {NoStop}%
\bibitem [{\citenamefont {Morel}\ and\ \citenamefont
  {Anderson}(1962)}]{Morel_1962}%
  \BibitemOpen
  \bibfield  {author} {\bibinfo {author} {\bibfnamefont {P.}~\bibnamefont
  {Morel}}\ and\ \bibinfo {author} {\bibfnamefont {P.~W.}\ \bibnamefont
  {Anderson}},\ }\href@noop {} {\bibfield  {journal} {\bibinfo  {journal}
  {Phys. Rev.}\ }\textbf {\bibinfo {volume} {125}},\ \bibinfo {pages} {1263}
  (\bibinfo {year} {1962})}\BibitemShut {NoStop}%
\bibitem [{\citenamefont {Bilbro}\ and\ \citenamefont
  {McMillan}(1976)}]{Bilbro_1976}%
  \BibitemOpen
  \bibfield  {author} {\bibinfo {author} {\bibfnamefont {G.}~\bibnamefont
  {Bilbro}}\ and\ \bibinfo {author} {\bibfnamefont {W.~L.}\ \bibnamefont
  {McMillan}},\ }\href@noop {} {\bibfield  {journal} {\bibinfo  {journal}
  {Phys. Rev. B}\ }\textbf {\bibinfo {volume} {14}},\ \bibinfo {pages} {1887}
  (\bibinfo {year} {1976})}\BibitemShut {NoStop}%
\bibitem [{\citenamefont {Chen}\ \emph {et~al.}(2017)\citenamefont {Chen},
  \citenamefont {Zhang}, \citenamefont {Zhang}, \citenamefont {Dong},\ and\
  \citenamefont {Wang}}]{Chen_2017}%
  \BibitemOpen
  \bibfield  {author} {\bibinfo {author} {\bibfnamefont {R.~Y.}\ \bibnamefont
  {Chen}}, \bibinfo {author} {\bibfnamefont {S.~J.}\ \bibnamefont {Zhang}},
  \bibinfo {author} {\bibfnamefont {M.~Y.}\ \bibnamefont {Zhang}}, \bibinfo
  {author} {\bibfnamefont {T.}~\bibnamefont {Dong}}, \ and\ \bibinfo {author}
  {\bibfnamefont {N.~L.}\ \bibnamefont {Wang}},\ }\href@noop {} {\bibfield
  {journal} {\bibinfo  {journal} {Phys. Rev. Lett.}\ }\textbf {\bibinfo
  {volume} {118}},\ \bibinfo {pages} {107402} (\bibinfo {year}
  {2017})}\BibitemShut {NoStop}%
\bibitem [{\citenamefont {Yanase}\ \emph {et~al.}(2003)\citenamefont {Yanase},
  \citenamefont {Jujo}, \citenamefont {Nomura}, \citenamefont {Ikeda},
  \citenamefont {Hotta},\ and\ \citenamefont {Yamada}}]{Yanase_2003}%
  \BibitemOpen
  \bibfield  {author} {\bibinfo {author} {\bibfnamefont {Y.}~\bibnamefont
  {Yanase}}, \bibinfo {author} {\bibfnamefont {T.}~\bibnamefont {Jujo}},
  \bibinfo {author} {\bibfnamefont {T.}~\bibnamefont {Nomura}}, \bibinfo
  {author} {\bibfnamefont {H.}~\bibnamefont {Ikeda}}, \bibinfo {author}
  {\bibfnamefont {T.}~\bibnamefont {Hotta}}, \ and\ \bibinfo {author}
  {\bibfnamefont {K.}~\bibnamefont {Yamada}},\ }\href@noop {} {\bibfield
  {journal} {\bibinfo  {journal} {Physics Reports}\ }\textbf {\bibinfo {volume}
  {387}},\ \bibinfo {pages} {1} (\bibinfo {year} {2003})}\BibitemShut {NoStop}%
\end{thebibliography}%


\begin{thebibliography}{21}%
\makeatletter
\providecommand \@ifxundefined [1]{%
 \@ifx{#1\undefined}
}%
\providecommand \@ifnum [1]{%
 \ifnum #1\expandafter \@firstoftwo
 \else \expandafter \@secondoftwo
 \fi
}%
\providecommand \@ifx [1]{%
 \ifx #1\expandafter \@firstoftwo
 \else \expandafter \@secondoftwo
 \fi
}%
\providecommand \natexlab [1]{#1}%
\providecommand \enquote  [1]{``#1''}%
\providecommand \bibnamefont  [1]{#1}%
\providecommand \bibfnamefont [1]{#1}%
\providecommand \citenamefont [1]{#1}%
\providecommand \href@noop [0]{\@secondoftwo}%
\providecommand \href [0]{\begingroup \@sanitize@url \@href}%
\providecommand \@href[1]{\@@startlink{#1}\@@href}%
\providecommand \@@href[1]{\endgroup#1\@@endlink}%
\providecommand \@sanitize@url [0]{\catcode `\\12\catcode `\$12\catcode
  `\&12\catcode `\#12\catcode `\^12\catcode `\_12\catcode `\%12\relax}%
\providecommand \@@startlink[1]{}%
\providecommand \@@endlink[0]{}%
\providecommand \url  [0]{\begingroup\@sanitize@url \@url }%
\providecommand \@url [1]{\endgroup\@href {#1}{\urlprefix }}%
\providecommand \urlprefix  [0]{URL }%
\providecommand \Eprint [0]{\href }%
\providecommand \doibase [0]{http://dx.doi.org/}%
\providecommand \selectlanguage [0]{\@gobble}%
\providecommand \bibinfo  [0]{\@secondoftwo}%
\providecommand \bibfield  [0]{\@secondoftwo}%
\providecommand \translation [1]{[#1]}%
\providecommand \BibitemOpen [0]{}%
\providecommand \bibitemStop [0]{}%
\providecommand \bibitemNoStop [0]{.\EOS\space}%
\providecommand \EOS [0]{\spacefactor3000\relax}%
\providecommand \BibitemShut  [1]{\csname bibitem#1\endcsname}%
\let\auto@bib@innerbib\@empty
\bibitem [{\citenamefont {Kittel}(2004)}]{Kittel}%
  \BibitemOpen
  \bibfield  {author} {\bibinfo {author} {\bibfnamefont {C.}~\bibnamefont
  {Kittel}},\ }\href@noop {} {\emph {\bibinfo {title} {Introduction to Solid
  State Physics}}},\ \bibinfo {edition} {8th}\ ed.\ (\bibinfo  {publisher}
  {John Wiley \& Sons},\ \bibinfo {year} {2004})\BibitemShut {NoStop}%
\bibitem [{\citenamefont {Giannozzi}\ \emph {et~al.}(2009)\citenamefont
  {Giannozzi}, \citenamefont {Baroni}, \citenamefont {Bonini}, \citenamefont
  {Calandra}, \citenamefont {Car}, \citenamefont {Cavazzoni}, \citenamefont
  {Ceresoli}, \citenamefont {Chiarotti}, \citenamefont {Cococcioni},
  \citenamefont {Dabo}, \citenamefont {Corso}, \citenamefont {de~Gironcoli},
  \citenamefont {Fabris}, \citenamefont {Fratesi}, \citenamefont {Gebauer},
  \citenamefont {Gerstmann}, \citenamefont {Gougoussis}, \citenamefont
  {Kokalj}, \citenamefont {Lazzeri}, \citenamefont {Martin-Samos},
  \citenamefont {Marzari}, \citenamefont {Mauri}, \citenamefont {Mazzarello},
  \citenamefont {Paolini}, \citenamefont {Pasquarello}, \citenamefont
  {Paulatto}, \citenamefont {Sbraccia}, \citenamefont {Scandolo}, \citenamefont
  {Sclauzero}, \citenamefont {Seitsonen}, \citenamefont {Smogunov},
  \citenamefont {Umari},\ and\ \citenamefont {Wentzcovitch}}]{Giannozzi_2009}%
  \BibitemOpen
  \bibfield  {author} {\bibinfo {author} {\bibfnamefont {P.}~\bibnamefont
  {Giannozzi}}, \bibinfo {author} {\bibfnamefont {S.}~\bibnamefont {Baroni}},
  \bibinfo {author} {\bibfnamefont {N.}~\bibnamefont {Bonini}}, \bibinfo
  {author} {\bibfnamefont {M.}~\bibnamefont {Calandra}}, \bibinfo {author}
  {\bibfnamefont {R.}~\bibnamefont {Car}}, \bibinfo {author} {\bibfnamefont
  {C.}~\bibnamefont {Cavazzoni}}, \bibinfo {author} {\bibfnamefont
  {D.}~\bibnamefont {Ceresoli}}, \bibinfo {author} {\bibfnamefont {G.~L.}\
  \bibnamefont {Chiarotti}}, \bibinfo {author} {\bibfnamefont {M.}~\bibnamefont
  {Cococcioni}}, \bibinfo {author} {\bibfnamefont {I.}~\bibnamefont {Dabo}},
  \bibinfo {author} {\bibfnamefont {A.~D.}\ \bibnamefont {Corso}}, \bibinfo
  {author} {\bibfnamefont {S.}~\bibnamefont {de~Gironcoli}}, \bibinfo {author}
  {\bibfnamefont {S.}~\bibnamefont {Fabris}}, \bibinfo {author} {\bibfnamefont
  {G.}~\bibnamefont {Fratesi}}, \bibinfo {author} {\bibfnamefont
  {R.}~\bibnamefont {Gebauer}}, \bibinfo {author} {\bibfnamefont
  {U.}~\bibnamefont {Gerstmann}}, \bibinfo {author} {\bibfnamefont
  {C.}~\bibnamefont {Gougoussis}}, \bibinfo {author} {\bibfnamefont
  {A.}~\bibnamefont {Kokalj}}, \bibinfo {author} {\bibfnamefont
  {M.}~\bibnamefont {Lazzeri}}, \bibinfo {author} {\bibfnamefont
  {L.}~\bibnamefont {Martin-Samos}}, \bibinfo {author} {\bibfnamefont
  {N.}~\bibnamefont {Marzari}}, \bibinfo {author} {\bibfnamefont
  {F.}~\bibnamefont {Mauri}}, \bibinfo {author} {\bibfnamefont
  {R.}~\bibnamefont {Mazzarello}}, \bibinfo {author} {\bibfnamefont
  {S.}~\bibnamefont {Paolini}}, \bibinfo {author} {\bibfnamefont
  {A.}~\bibnamefont {Pasquarello}}, \bibinfo {author} {\bibfnamefont
  {L.}~\bibnamefont {Paulatto}}, \bibinfo {author} {\bibfnamefont
  {C.}~\bibnamefont {Sbraccia}}, \bibinfo {author} {\bibfnamefont
  {S.}~\bibnamefont {Scandolo}}, \bibinfo {author} {\bibfnamefont
  {G.}~\bibnamefont {Sclauzero}}, \bibinfo {author} {\bibfnamefont {A.~P.}\
  \bibnamefont {Seitsonen}}, \bibinfo {author} {\bibfnamefont {A.}~\bibnamefont
  {Smogunov}}, \bibinfo {author} {\bibfnamefont {P.}~\bibnamefont {Umari}}, \
  and\ \bibinfo {author} {\bibfnamefont {R.~M.}\ \bibnamefont {Wentzcovitch}},\
  }\href@noop {} {\bibfield  {journal} {\bibinfo  {journal} {J. Phys.: Condens.
  Matter}\ }\textbf {\bibinfo {volume} {21}},\ \bibinfo {pages} {395502}
  (\bibinfo {year} {2009})}\BibitemShut {NoStop}%
\bibitem [{\citenamefont {Giannozzi}\ \emph {et~al.}(2017)\citenamefont
  {Giannozzi}, \citenamefont {Andreussi}, \citenamefont {Brumme}, \citenamefont
  {Bunau}, \citenamefont {{Buongiorno Nardelli}}, \citenamefont {Calandra},
  \citenamefont {Car}, \citenamefont {Cavazzoni}, \citenamefont {Ceresoli},
  \citenamefont {Cococcioni}, \citenamefont {Colonna}, \citenamefont
  {Carnimeo}, \citenamefont {{Dal Corso}}, \citenamefont {{De Gironcoli}},
  \citenamefont {Delugas}, \citenamefont {Distasio}, \citenamefont {Ferretti},
  \citenamefont {Floris}, \citenamefont {Fratesi}, \citenamefont {Fugallo},
  \citenamefont {Gebauer}, \citenamefont {Gerstmann}, \citenamefont {Giustino},
  \citenamefont {Gorni}, \citenamefont {Jia}, \citenamefont {Kawamura},
  \citenamefont {Ko}, \citenamefont {Kokalj}, \citenamefont
  {K{\"u}c{\"u}kbenli}, \citenamefont {Lazzeri}, \citenamefont {Marsili},
  \citenamefont {Marzari}, \citenamefont {Mauri}, \citenamefont {Nguyen},
  \citenamefont {Nguyen}, \citenamefont {Otero-De-La-Roza}, \citenamefont
  {Paulatto}, \citenamefont {Ponc{\'e}}, \citenamefont {Rocca}, \citenamefont
  {Sabatini}, \citenamefont {Santra}, \citenamefont {Schlipf}, \citenamefont
  {Seitsonen}, \citenamefont {Smogunov}, \citenamefont {Timrov}, \citenamefont
  {Thonhauser}, \citenamefont {Umari}, \citenamefont {Vast}, \citenamefont
  {Wu},\ and\ \citenamefont {Baroni}}]{Giannozzi_2017}%
  \BibitemOpen
  \bibfield  {author} {\bibinfo {author} {\bibfnamefont {P.}~\bibnamefont
  {Giannozzi}}, \bibinfo {author} {\bibfnamefont {O.}~\bibnamefont
  {Andreussi}}, \bibinfo {author} {\bibfnamefont {T.}~\bibnamefont {Brumme}},
  \bibinfo {author} {\bibfnamefont {O.}~\bibnamefont {Bunau}}, \bibinfo
  {author} {\bibfnamefont {M.}~\bibnamefont {{Buongiorno Nardelli}}}, \bibinfo
  {author} {\bibfnamefont {M.}~\bibnamefont {Calandra}}, \bibinfo {author}
  {\bibfnamefont {R.}~\bibnamefont {Car}}, \bibinfo {author} {\bibfnamefont
  {C.}~\bibnamefont {Cavazzoni}}, \bibinfo {author} {\bibfnamefont
  {D.}~\bibnamefont {Ceresoli}}, \bibinfo {author} {\bibfnamefont
  {M.}~\bibnamefont {Cococcioni}}, \bibinfo {author} {\bibfnamefont
  {N.}~\bibnamefont {Colonna}}, \bibinfo {author} {\bibfnamefont
  {I.}~\bibnamefont {Carnimeo}}, \bibinfo {author} {\bibfnamefont
  {A.}~\bibnamefont {{Dal Corso}}}, \bibinfo {author} {\bibfnamefont
  {S.}~\bibnamefont {{De Gironcoli}}}, \bibinfo {author} {\bibfnamefont
  {P.}~\bibnamefont {Delugas}}, \bibinfo {author} {\bibfnamefont
  {R.}~\bibnamefont {Distasio}}, \bibinfo {author} {\bibfnamefont
  {A.}~\bibnamefont {Ferretti}}, \bibinfo {author} {\bibfnamefont
  {A.}~\bibnamefont {Floris}}, \bibinfo {author} {\bibfnamefont
  {G.}~\bibnamefont {Fratesi}}, \bibinfo {author} {\bibfnamefont
  {G.}~\bibnamefont {Fugallo}}, \bibinfo {author} {\bibfnamefont
  {R.}~\bibnamefont {Gebauer}}, \bibinfo {author} {\bibfnamefont
  {U.}~\bibnamefont {Gerstmann}}, \bibinfo {author} {\bibfnamefont
  {F.}~\bibnamefont {Giustino}}, \bibinfo {author} {\bibfnamefont
  {T.}~\bibnamefont {Gorni}}, \bibinfo {author} {\bibfnamefont
  {J.}~\bibnamefont {Jia}}, \bibinfo {author} {\bibfnamefont {M.}~\bibnamefont
  {Kawamura}}, \bibinfo {author} {\bibfnamefont {H.}~\bibnamefont {Ko}},
  \bibinfo {author} {\bibfnamefont {A.}~\bibnamefont {Kokalj}}, \bibinfo
  {author} {\bibfnamefont {E.}~\bibnamefont {K{\"u}c{\"u}kbenli}}, \bibinfo
  {author} {\bibfnamefont {M.}~\bibnamefont {Lazzeri}}, \bibinfo {author}
  {\bibfnamefont {M.}~\bibnamefont {Marsili}}, \bibinfo {author} {\bibfnamefont
  {N.}~\bibnamefont {Marzari}}, \bibinfo {author} {\bibfnamefont
  {F.}~\bibnamefont {Mauri}}, \bibinfo {author} {\bibfnamefont
  {N.}~\bibnamefont {Nguyen}}, \bibinfo {author} {\bibfnamefont
  {H.}~\bibnamefont {Nguyen}}, \bibinfo {author} {\bibfnamefont
  {A.}~\bibnamefont {Otero-De-La-Roza}}, \bibinfo {author} {\bibfnamefont
  {L.}~\bibnamefont {Paulatto}}, \bibinfo {author} {\bibfnamefont
  {S.}~\bibnamefont {Ponc{\'e}}}, \bibinfo {author} {\bibfnamefont
  {D.}~\bibnamefont {Rocca}}, \bibinfo {author} {\bibfnamefont
  {R.}~\bibnamefont {Sabatini}}, \bibinfo {author} {\bibfnamefont
  {B.}~\bibnamefont {Santra}}, \bibinfo {author} {\bibfnamefont
  {M.}~\bibnamefont {Schlipf}}, \bibinfo {author} {\bibfnamefont
  {A.}~\bibnamefont {Seitsonen}}, \bibinfo {author} {\bibfnamefont
  {A.}~\bibnamefont {Smogunov}}, \bibinfo {author} {\bibfnamefont
  {I.}~\bibnamefont {Timrov}}, \bibinfo {author} {\bibfnamefont
  {T.}~\bibnamefont {Thonhauser}}, \bibinfo {author} {\bibfnamefont
  {P.}~\bibnamefont {Umari}}, \bibinfo {author} {\bibfnamefont
  {N.}~\bibnamefont {Vast}}, \bibinfo {author} {\bibfnamefont {X.}~\bibnamefont
  {Wu}}, \ and\ \bibinfo {author} {\bibfnamefont {S.}~\bibnamefont {Baroni}},\
  }\href@noop {} {\bibfield  {journal} {\bibinfo  {journal} {J. Phys.: Condens.
  Matter}\ }\textbf {\bibinfo {volume} {29}},\ \bibinfo {pages} {465901}
  (\bibinfo {year} {2017})}\BibitemShut {NoStop}%
\bibitem [{pse()}]{pseudo}%
  \BibitemOpen
  \href@noop {} {}\bibinfo {note} {We used
  \url{La.pbe-spfn-rrkjus_psl.1.0.0.UPF} for La,
  \url{Ag.pbe-n-rrkjus_psl.1.0.0.UPF} for Ag, and
  \url{Sb.pbe-n-rrkjus_psl.1.0.0.UPF} for Sb in \url{PSlibrary}.}\BibitemShut
  {Stop}%
\bibitem [{\citenamefont {Akiba}\ \emph {et~al.}(2021)\citenamefont {Akiba},
  \citenamefont {Nishimori}, \citenamefont {Umeshita},\ and\ \citenamefont
  {Kobayashi}}]{Akiba_2021}%
  \BibitemOpen
  \bibfield  {author} {\bibinfo {author} {\bibfnamefont {K.}~\bibnamefont
  {Akiba}}, \bibinfo {author} {\bibfnamefont {H.}~\bibnamefont {Nishimori}},
  \bibinfo {author} {\bibfnamefont {N.}~\bibnamefont {Umeshita}}, \ and\
  \bibinfo {author} {\bibfnamefont {T.~C.}\ \bibnamefont {Kobayashi}},\
  }\href@noop {} {\bibfield  {journal} {\bibinfo  {journal} {Phys. Rev. B}\
  }\textbf {\bibinfo {volume} {103}},\ \bibinfo {pages} {085134} (\bibinfo
  {year} {2021})}\BibitemShut {NoStop}%
\bibitem [{\citenamefont {Bud'ko}\ \emph {et~al.}(2006)\citenamefont {Bud'ko},
  \citenamefont {Wiener}, \citenamefont {Ribeiro}, \citenamefont {Canfield},
  \citenamefont {Lee}, \citenamefont {Vogt},\ and\ \citenamefont
  {Lacerda}}]{Budko_2006}%
  \BibitemOpen
  \bibfield  {author} {\bibinfo {author} {\bibfnamefont {S.~L.}\ \bibnamefont
  {Bud'ko}}, \bibinfo {author} {\bibfnamefont {T.~A.}\ \bibnamefont {Wiener}},
  \bibinfo {author} {\bibfnamefont {R.~A.}\ \bibnamefont {Ribeiro}}, \bibinfo
  {author} {\bibfnamefont {P.~C.}\ \bibnamefont {Canfield}}, \bibinfo {author}
  {\bibfnamefont {Y.}~\bibnamefont {Lee}}, \bibinfo {author} {\bibfnamefont
  {T.}~\bibnamefont {Vogt}}, \ and\ \bibinfo {author} {\bibfnamefont {A.~H.}\
  \bibnamefont {Lacerda}},\ }\href@noop {} {\bibfield  {journal} {\bibinfo
  {journal} {Phys. Rev. B}\ }\textbf {\bibinfo {volume} {73}},\ \bibinfo
  {pages} {184111} (\bibinfo {year} {2006})}\BibitemShut {NoStop}%
\bibitem [{\citenamefont {Singha}\ \emph {et~al.}(2020)\citenamefont {Singha},
  \citenamefont {Samanta}, \citenamefont {Bhattacharya}, \citenamefont
  {Chatterjee}, \citenamefont {Roy}, \citenamefont {Wang}, \citenamefont
  {Singha},\ and\ \citenamefont {Mandal}}]{Singha_2020}%
  \BibitemOpen
  \bibfield  {author} {\bibinfo {author} {\bibfnamefont {R.}~\bibnamefont
  {Singha}}, \bibinfo {author} {\bibfnamefont {S.}~\bibnamefont {Samanta}},
  \bibinfo {author} {\bibfnamefont {T.~S.}\ \bibnamefont {Bhattacharya}},
  \bibinfo {author} {\bibfnamefont {S.}~\bibnamefont {Chatterjee}}, \bibinfo
  {author} {\bibfnamefont {S.}~\bibnamefont {Roy}}, \bibinfo {author}
  {\bibfnamefont {L.}~\bibnamefont {Wang}}, \bibinfo {author} {\bibfnamefont
  {A.}~\bibnamefont {Singha}}, \ and\ \bibinfo {author} {\bibfnamefont
  {P.}~\bibnamefont {Mandal}},\ }\href@noop {} {\bibfield  {journal} {\bibinfo
  {journal} {Phys. Rev. B}\ }\textbf {\bibinfo {volume} {102}},\ \bibinfo
  {pages} {205103} (\bibinfo {year} {2020})}\BibitemShut {NoStop}%
\bibitem [{\citenamefont {Kawamura}\ \emph {et~al.}(2014)\citenamefont
  {Kawamura}, \citenamefont {Gohda},\ and\ \citenamefont
  {Tsuneyuki}}]{Kawamura_2014}%
  \BibitemOpen
  \bibfield  {author} {\bibinfo {author} {\bibfnamefont {M.}~\bibnamefont
  {Kawamura}}, \bibinfo {author} {\bibfnamefont {Y.}~\bibnamefont {Gohda}}, \
  and\ \bibinfo {author} {\bibfnamefont {S.}~\bibnamefont {Tsuneyuki}},\
  }\href@noop {} {\bibfield  {journal} {\bibinfo  {journal} {Phys. Rev. B}\
  }\textbf {\bibinfo {volume} {89}},\ \bibinfo {pages} {094515} (\bibinfo
  {year} {2014})}\BibitemShut {NoStop}%
\bibitem [{\citenamefont {Ruszala}\ \emph {et~al.}(2020)\citenamefont
  {Ruszala}, \citenamefont {Winiarski},\ and\ \citenamefont
  {Samsel-Czekala}}]{Ruszala_2020}%
  \BibitemOpen
  \bibfield  {author} {\bibinfo {author} {\bibfnamefont {P.}~\bibnamefont
  {Ruszala}}, \bibinfo {author} {\bibfnamefont {M.}~\bibnamefont {Winiarski}},
  \ and\ \bibinfo {author} {\bibfnamefont {M.}~\bibnamefont {Samsel-Czekala}},\
  }\href@noop {} {\bibfield  {journal} {\bibinfo  {journal} {Acta Physica
  Polonica A}\ }\textbf {\bibinfo {volume} {138}},\ \bibinfo {pages} {748}
  (\bibinfo {year} {2020})}\BibitemShut {NoStop}%
\bibitem [{\citenamefont {Akiba}\ \emph {et~al.}(2022)\citenamefont {Akiba},
  \citenamefont {Umeshita},\ and\ \citenamefont {Kobayashi}}]{Akiba_2022}%
  \BibitemOpen
  \bibfield  {author} {\bibinfo {author} {\bibfnamefont {K.}~\bibnamefont
  {Akiba}}, \bibinfo {author} {\bibfnamefont {N.}~\bibnamefont {Umeshita}}, \
  and\ \bibinfo {author} {\bibfnamefont {T.~C.}\ \bibnamefont {Kobayashi}},\
  }\href@noop {} {\bibfield  {journal} {\bibinfo  {journal} {Phys. Rev. B}\
  }\textbf {\bibinfo {volume} {105}},\ \bibinfo {pages} {035108} (\bibinfo
  {year} {2022})}\BibitemShut {NoStop}%
\bibitem [{\citenamefont {Ziman}(2001)}]{Ziman_ep}%
  \BibitemOpen
  \bibfield  {author} {\bibinfo {author} {\bibfnamefont {J.~M.}\ \bibnamefont
  {Ziman}},\ }\href@noop {} {\emph {\bibinfo {title} {Electrons and Phonons:
  The Theory of Transport Phenomena in Solids}}}\ (\bibinfo  {publisher}
  {Oxford University Press},\ \bibinfo {year} {2001})\BibitemShut {NoStop}%
\bibitem [{\citenamefont {Takeuchi}\ \emph {et~al.}(2003)\citenamefont
  {Takeuchi}, \citenamefont {Thamizhavel}, \citenamefont {Okubo}, \citenamefont
  {Yamada}, \citenamefont {Nakamura}, \citenamefont {Yamamoto}, \citenamefont
  {Inada}, \citenamefont {Sugiyama}, \citenamefont {Galatanu}, \citenamefont
  {Yamamoto}, \citenamefont {Kindo}, \citenamefont {Ebihara},\ and\
  \citenamefont {\ifmmode~\bar{O}\else \={O}\fi{}nuki}}]{Takeuchi_2003}%
  \BibitemOpen
  \bibfield  {author} {\bibinfo {author} {\bibfnamefont {T.}~\bibnamefont
  {Takeuchi}}, \bibinfo {author} {\bibfnamefont {A.}~\bibnamefont
  {Thamizhavel}}, \bibinfo {author} {\bibfnamefont {T.}~\bibnamefont {Okubo}},
  \bibinfo {author} {\bibfnamefont {M.}~\bibnamefont {Yamada}}, \bibinfo
  {author} {\bibfnamefont {N.}~\bibnamefont {Nakamura}}, \bibinfo {author}
  {\bibfnamefont {T.}~\bibnamefont {Yamamoto}}, \bibinfo {author}
  {\bibfnamefont {Y.}~\bibnamefont {Inada}}, \bibinfo {author} {\bibfnamefont
  {K.}~\bibnamefont {Sugiyama}}, \bibinfo {author} {\bibfnamefont
  {A.}~\bibnamefont {Galatanu}}, \bibinfo {author} {\bibfnamefont
  {E.}~\bibnamefont {Yamamoto}}, \bibinfo {author} {\bibfnamefont
  {K.}~\bibnamefont {Kindo}}, \bibinfo {author} {\bibfnamefont
  {T.}~\bibnamefont {Ebihara}}, \ and\ \bibinfo {author} {\bibfnamefont
  {Y.}~\bibnamefont {\ifmmode~\bar{O}\else \={O}\fi{}nuki}},\ }\href@noop {}
  {\bibfield  {journal} {\bibinfo  {journal} {Phys. Rev. B}\ }\textbf {\bibinfo
  {volume} {67}},\ \bibinfo {pages} {064403} (\bibinfo {year}
  {2003})}\BibitemShut {NoStop}%
\bibitem [{\citenamefont {Pizzi}\ \emph {et~al.}(2020)\citenamefont {Pizzi},
  \citenamefont {Vitale}, \citenamefont {Arita}, \citenamefont {Bluegel},
  \citenamefont {Freimuth}, \citenamefont {G{\'e}ranton}, \citenamefont
  {Gibertini}, \citenamefont {Gresch}, \citenamefont {Johnson}, \citenamefont
  {Koretsune}, \citenamefont {Ibanez}, \citenamefont {Lee}, \citenamefont
  {Lihm}, \citenamefont {Marchand}, \citenamefont {Marrazzo}, \citenamefont
  {Mokrousov}, \citenamefont {Mustafa}, \citenamefont {Nohara}, \citenamefont
  {Nomura}, \citenamefont {Paulatto}, \citenamefont {Ponc{\'e}}, \citenamefont
  {Ponweiser}, \citenamefont {Qiao}, \citenamefont {Th{\"o}le}, \citenamefont
  {Tsirkin}, \citenamefont {Wierzbowska}, \citenamefont {Marzari},
  \citenamefont {Vanderbilt}, \citenamefont {Souza}, \citenamefont {Mostofi},\
  and\ \citenamefont {Yates}}]{Pizzi_2020}%
  \BibitemOpen
  \bibfield  {author} {\bibinfo {author} {\bibfnamefont {G.}~\bibnamefont
  {Pizzi}}, \bibinfo {author} {\bibfnamefont {V.}~\bibnamefont {Vitale}},
  \bibinfo {author} {\bibfnamefont {R.}~\bibnamefont {Arita}}, \bibinfo
  {author} {\bibfnamefont {S.}~\bibnamefont {Bluegel}}, \bibinfo {author}
  {\bibfnamefont {F.}~\bibnamefont {Freimuth}}, \bibinfo {author}
  {\bibfnamefont {G.}~\bibnamefont {G{\'e}ranton}}, \bibinfo {author}
  {\bibfnamefont {M.}~\bibnamefont {Gibertini}}, \bibinfo {author}
  {\bibfnamefont {D.}~\bibnamefont {Gresch}}, \bibinfo {author} {\bibfnamefont
  {C.}~\bibnamefont {Johnson}}, \bibinfo {author} {\bibfnamefont
  {T.}~\bibnamefont {Koretsune}}, \bibinfo {author} {\bibfnamefont
  {J.}~\bibnamefont {Ibanez}}, \bibinfo {author} {\bibfnamefont
  {H.}~\bibnamefont {Lee}}, \bibinfo {author} {\bibfnamefont {J.-M.}\
  \bibnamefont {Lihm}}, \bibinfo {author} {\bibfnamefont {D.}~\bibnamefont
  {Marchand}}, \bibinfo {author} {\bibfnamefont {A.}~\bibnamefont {Marrazzo}},
  \bibinfo {author} {\bibfnamefont {Y.}~\bibnamefont {Mokrousov}}, \bibinfo
  {author} {\bibfnamefont {J.~I.}\ \bibnamefont {Mustafa}}, \bibinfo {author}
  {\bibfnamefont {Y.}~\bibnamefont {Nohara}}, \bibinfo {author} {\bibfnamefont
  {Y.}~\bibnamefont {Nomura}}, \bibinfo {author} {\bibfnamefont
  {L.}~\bibnamefont {Paulatto}}, \bibinfo {author} {\bibfnamefont
  {S.}~\bibnamefont {Ponc{\'e}}}, \bibinfo {author} {\bibfnamefont
  {T.}~\bibnamefont {Ponweiser}}, \bibinfo {author} {\bibfnamefont
  {J.}~\bibnamefont {Qiao}}, \bibinfo {author} {\bibfnamefont {F.}~\bibnamefont
  {Th{\"o}le}}, \bibinfo {author} {\bibfnamefont {S.~S.}\ \bibnamefont
  {Tsirkin}}, \bibinfo {author} {\bibfnamefont {M.}~\bibnamefont
  {Wierzbowska}}, \bibinfo {author} {\bibfnamefont {N.}~\bibnamefont
  {Marzari}}, \bibinfo {author} {\bibfnamefont {D.}~\bibnamefont {Vanderbilt}},
  \bibinfo {author} {\bibfnamefont {I.}~\bibnamefont {Souza}}, \bibinfo
  {author} {\bibfnamefont {A.~A.}\ \bibnamefont {Mostofi}}, \ and\ \bibinfo
  {author} {\bibfnamefont {J.~R.}\ \bibnamefont {Yates}},\ }\href@noop {}
  {\bibfield  {journal} {\bibinfo  {journal} {J. Phys.: Condens. Matter}\
  }\textbf {\bibinfo {volume} {32}},\ \bibinfo {pages} {165902} (\bibinfo
  {year} {2020})}\BibitemShut {NoStop}%
\bibitem [{\citenamefont {Ponc\'e}\ \emph {et~al.}(2016)\citenamefont
  {Ponc\'e}, \citenamefont {Margine}, \citenamefont {Verdi},\ and\
  \citenamefont {Giustino}}]{Ponce_2016}%
  \BibitemOpen
  \bibfield  {author} {\bibinfo {author} {\bibfnamefont {S.}~\bibnamefont
  {Ponc\'e}}, \bibinfo {author} {\bibfnamefont {E.}~\bibnamefont {Margine}},
  \bibinfo {author} {\bibfnamefont {C.}~\bibnamefont {Verdi}}, \ and\ \bibinfo
  {author} {\bibfnamefont {F.}~\bibnamefont {Giustino}},\ }\href@noop {}
  {\bibfield  {journal} {\bibinfo  {journal} {Comput. Phys. Commun.}\ }\textbf
  {\bibinfo {volume} {209}},\ \bibinfo {pages} {116} (\bibinfo {year}
  {2016})}\BibitemShut {NoStop}%
\bibitem [{\citenamefont {McMillan}(1968)}]{McMillan_1968}%
  \BibitemOpen
  \bibfield  {author} {\bibinfo {author} {\bibfnamefont {W.~L.}\ \bibnamefont
  {McMillan}},\ }\href@noop {} {\bibfield  {journal} {\bibinfo  {journal}
  {Phys. Rev.}\ }\textbf {\bibinfo {volume} {167}},\ \bibinfo {pages} {331}
  (\bibinfo {year} {1968})}\BibitemShut {NoStop}%
\bibitem [{\citenamefont {Dynes}(1972)}]{Dynes_1972}%
  \BibitemOpen
  \bibfield  {author} {\bibinfo {author} {\bibfnamefont {R.~C.}\ \bibnamefont
  {Dynes}},\ }\href@noop {} {\bibfield  {journal} {\bibinfo  {journal} {Solid
  State Commun.}\ }\textbf {\bibinfo {volume} {10}},\ \bibinfo {pages} {615}
  (\bibinfo {year} {1972})}\BibitemShut {NoStop}%
\bibitem [{\citenamefont {Allen}\ and\ \citenamefont
  {Dynes}(1975)}]{Allen_1975}%
  \BibitemOpen
  \bibfield  {author} {\bibinfo {author} {\bibfnamefont {P.~B.}\ \bibnamefont
  {Allen}}\ and\ \bibinfo {author} {\bibfnamefont {R.~C.}\ \bibnamefont
  {Dynes}},\ }\href@noop {} {\bibfield  {journal} {\bibinfo  {journal} {Phys.
  Rev. B}\ }\textbf {\bibinfo {volume} {12}},\ \bibinfo {pages} {905} (\bibinfo
  {year} {1975})}\BibitemShut {NoStop}%
\bibitem [{\citenamefont {Morel}\ and\ \citenamefont
  {Anderson}(1962)}]{Morel_1962}%
  \BibitemOpen
  \bibfield  {author} {\bibinfo {author} {\bibfnamefont {P.}~\bibnamefont
  {Morel}}\ and\ \bibinfo {author} {\bibfnamefont {P.~W.}\ \bibnamefont
  {Anderson}},\ }\href@noop {} {\bibfield  {journal} {\bibinfo  {journal}
  {Phys. Rev.}\ }\textbf {\bibinfo {volume} {125}},\ \bibinfo {pages} {1263}
  (\bibinfo {year} {1962})}\BibitemShut {NoStop}%
\bibitem [{\citenamefont {Giustino}\ \emph {et~al.}(2007)\citenamefont
  {Giustino}, \citenamefont {Cohen},\ and\ \citenamefont
  {Louie}}]{Giustino_2007}%
  \BibitemOpen
  \bibfield  {author} {\bibinfo {author} {\bibfnamefont {F.}~\bibnamefont
  {Giustino}}, \bibinfo {author} {\bibfnamefont {M.~L.}\ \bibnamefont {Cohen}},
  \ and\ \bibinfo {author} {\bibfnamefont {S.~G.}\ \bibnamefont {Louie}},\
  }\href@noop {} {\bibfield  {journal} {\bibinfo  {journal} {Phys. Rev. B}\
  }\textbf {\bibinfo {volume} {76}},\ \bibinfo {pages} {165108} (\bibinfo
  {year} {2007})}\BibitemShut {NoStop}%
\bibitem [{con()}]{conv}%
  \BibitemOpen
  \href@noop {} {}\bibinfo {note} {We also confirmed that the calculations
  converge to almost identical $\lambda$ using quasi-random $25^3$ $k$- and
  $q$-grids.}\BibitemShut {Stop}%
\bibitem [{\citenamefont {Kawamura}(2019)}]{Kawamura_2019}%
  \BibitemOpen
  \bibfield  {author} {\bibinfo {author} {\bibfnamefont {M.}~\bibnamefont
  {Kawamura}},\ }\href@noop {} {\bibfield  {journal} {\bibinfo  {journal}
  {Comput. Phys. Commun.}\ }\textbf {\bibinfo {volume} {239}},\ \bibinfo
  {pages} {197} (\bibinfo {year} {2019})}\BibitemShut {NoStop}%
\end{thebibliography}%

\clearpage

\end{document}


\setcounter{equation}{0}
\setcounter{figure}{0}
\setcounter{table}{0}
\setcounter{page}{1}
\makeatletter
\renewcommand{\theequation}{S\arabic{equation}}
\renewcommand{\thefigure}{S\arabic{figure}}
\renewcommand{\thetable}{S\arabic{table}}

\preprint{APS/123-QED}

\title{Supplemental material for\\ ''Observation of superconductivity and its enhancement at the charge density wave critical point in LaAgSb$_2$''}

\author{Kazuto~Akiba}
\email{akb@okayama-u.ac.jp}
\affiliation{
Graduate School of Natural Science and Technology, 
Okayama University, Okayama 700-8530, Japan
}

\author{Nobuaki~Umeshita}
\affiliation{
Graduate School of Natural Science and Technology, 
Okayama University, Okayama 700-8530, Japan
}

\author{Tatsuo~C.~Kobayashi}
\affiliation{
Graduate School of Natural Science and Technology, 
Okayama University, Okayama 700-8530, Japan
}

\date{\today}
\begin{abstract}
This Supplemental material includes 6 chapters, Figs. S1 to S13, and Table S1 to S2.
\end{abstract}

\maketitle


\section{Magnetization measurements}

\begin{figure*}[]
\centering
\includegraphics[]{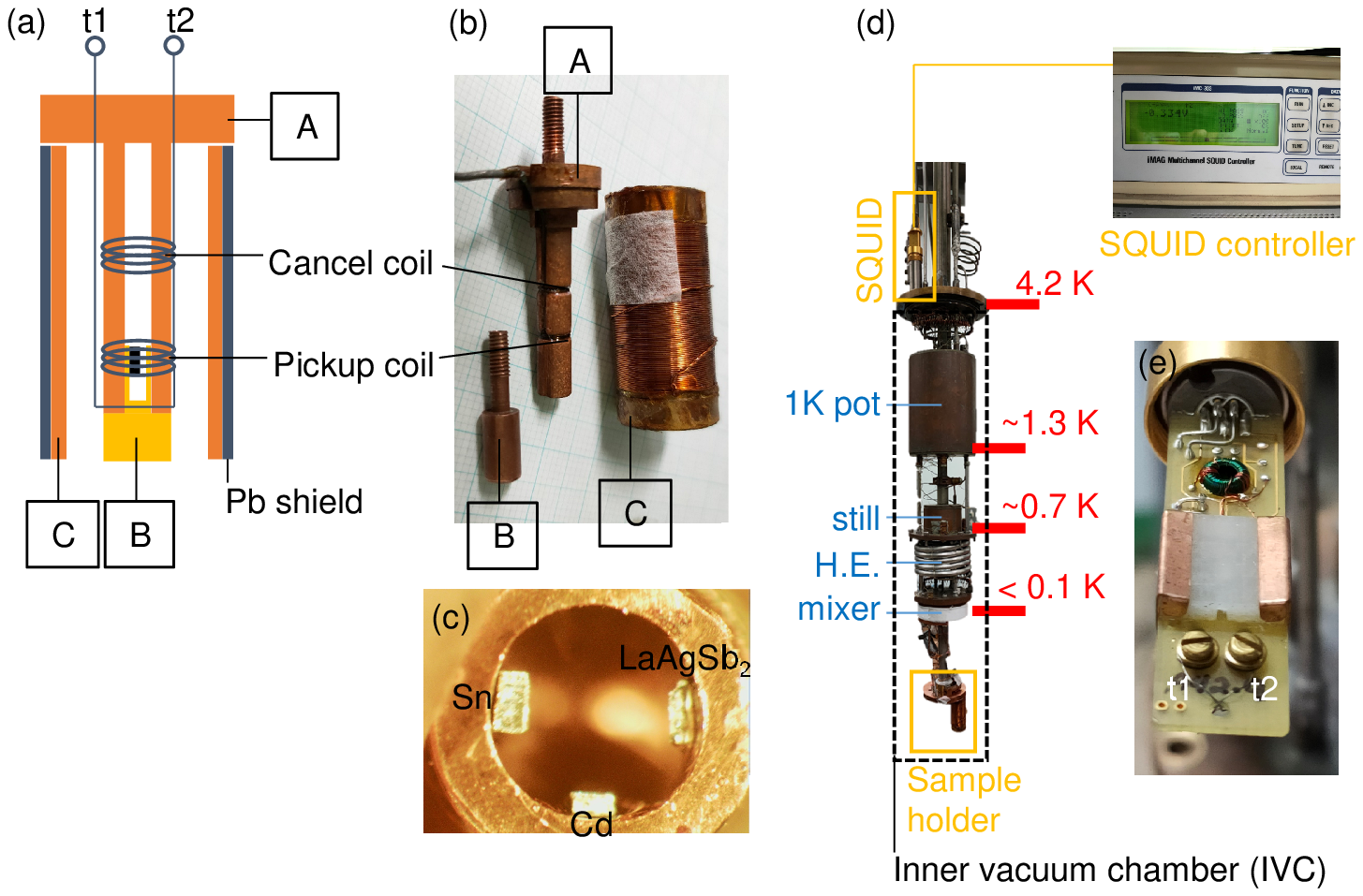}
\caption{
(a) Schematic of the signal-pickup system for magnetization measurement.
(b) Picture of the signal-pickup system. Parts A, B, and C correspond to those of (a).
(c) Picture of the samples set inside part B.
(d) Magnetization measurement system constructed on the dilution refrigerator.
(e) Picture of the DC SQUID element utilized in this study.
\label{figs1}}
\end{figure*}

We describe the magnetization measurement system utilized in this study.
Figures \ref{figs1}(a) and \ref{figs1}(b) show the schematic and actual picture of the measurement apparatus.
All parts are made of copper.
Part A has a pickup coil to detect the change in magnetic flux inside and cancel coil to reduce the background. These coils are wound by a superconducting NbTi wire.
Part B can be screwed into part A and has a cylindrical sample space at the center.
Samples are glued on the inner wall using Apiezon N grease ( Fig. \ref{figs1}(c)).
Part C is surrounded by a thin sheet of superconducting Pb to prevent the penetration of electronic noise from the outside.
Although part C has a solenoid coil on the outermost layer to apply an external magnetic field,
it has not been used here.
These parts are set on the sample holder of the dilution refrigerator (Fig. \ref{figs1}(d)).
The NbTi wire that starts from part A is shielded by a superconducting Nb tube
and is taken out from the inner vacuum chamber (IVC) of the refrigerator.
The ends of the wire are tightened on the terminals of the DC SQUID element (labeled as t1 and t2 in Fig. \ref{figs1}(e)) that is manufactured by Tristan Technologies, Inc.
The DC SQUID element is stored in a He bath, and hence its temperature is always held at 4.2 K.
The DC SQUID element is connected with a controller via a communication cable.
We can obtain the voltage signal on the controller, which is proportional to the change in the magnetic flux inside the pickup coil.
We also note that we did not intentionally apply an external magnetic field.
The situation corresponds to ``field cooling in a residual geomagnetic field''.

\begin{figure}[]
\centering
\includegraphics[]{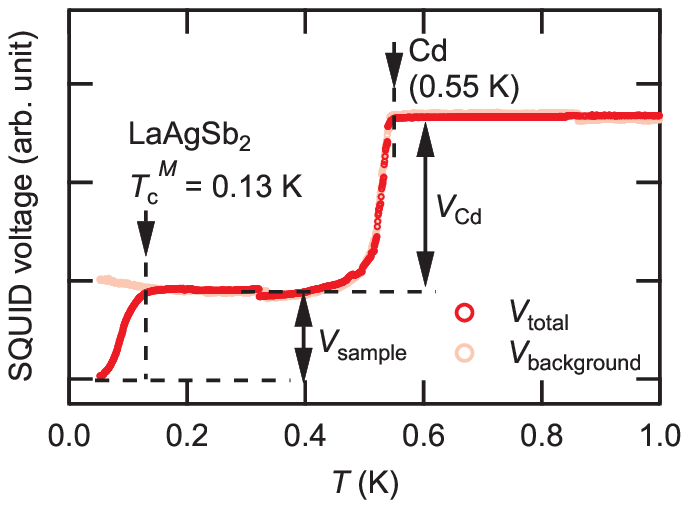}
\caption{
Temperature dependence of the SQUID voltage below 1 K.
$V_{total}$ (red markers) indicates the data obtained by placing the LaAgSb$_2$ sample and Cd (reference) in the pick-up coil.
$V_{background}$ (pink markers) indicates the data obtained by removing the sample from the above setup.
The anomaly observed at 0.55 K is due to the Meissner effect of Cd.
\label{figs2}}
\end{figure}

Figure \ref{figs2} shows the raw data to deduce the magnetization $M$ of the sample.
Note that we put a piece of Cd as a reference together with the LaAgSb$_2$ sample.
Cd is known to be an elemental superconductor with a transition temperature of 0.56 K \cite{Kittel}.
We measure the data with ($V_{total}$) and without ($V_{background}$) the sample.
Upon lowering the temperature, we observe a sharp anomaly at 0.55 K in both $V_{total}$ and $V_{background}$,
which is almost identical to the known superconducting transition temperature of Cd.
Upon further cooling, we can observe another anomaly that is found only in $V_{total}$, indicating the intrinsic magnetization response of the sample.
We obtained the data shown in the main text by subtracting $V_{background}$ from $V_{total}$.

Using the data presented in Fig. \ref{figs2}, we can estimate the superconducting volume fraction.
Here, we assume that the superconducting volume fraction of Cd is 100 \%.
Using the known volume of the sample ($1.3\times 10^{-4}$ cm$^3$), Cd ($1.0\times 10^{-4}$ cm$^3$), and the signal ratio of $V_{sample}:V_{Cd}=0.506:1$, where $V_{sample}$ and $V_{Cd}$ are defined in Fig. \ref{figs2},
the volume fraction of LaAgSb$_2$ is estimated to be 39 \% at 60 mK.

\clearpage

\section{Specific heat measurements}

\begin{figure*}[]
\centering
\includegraphics[]{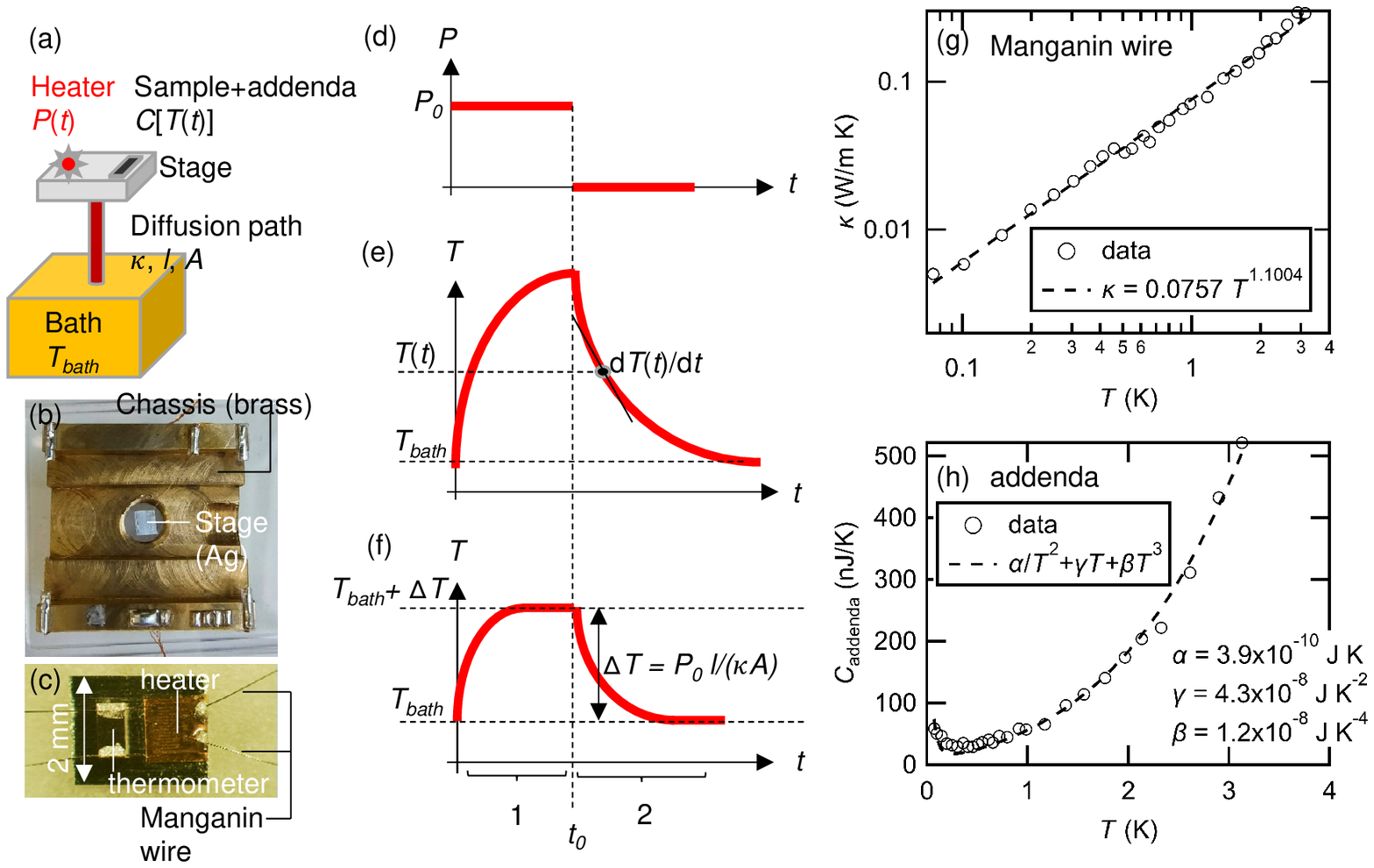}
\caption{
(a) Schematic of the heat capacity measurement cell.
Pictures of the (b) heat capacity measurement cell and (c) back of the sample stage.
(d) Typical heater power applied to the stage as a function of time.
(e) Typical temperature response measured by the thermometer in the sweep method.
(f) Typical temperature response measured by the thermometer in the step method.
(g) Temperature dependence of the thermal conductivity of the manganin wire.
(h) Temperature dependence of the addenda heat capacity.
\label{figs3}}
\end{figure*}

We employed the relaxation method for the specific heat capacity measurements at low temperature.
Figure \ref{figs3}(a) shows a schematic of the measurement cell.
Figures \ref{figs3}(b) and \ref{figs3}(c) present the whole view of the measurement cell and a
magnified view of the back of the sample stage, respectively.
We used a piece of Ag ribbon or Si single crystal (The Nilaco Corporation) as a sample stage.
The sample stage has a thermometer and a heater at the back 
as shown in Fig. \ref{figs3}(c).
We utilized a commercial chip resistor (RK73B1ETTP202J, KOA Corporation) and
a strain gauge (SKF-5414, Kyowa Electronic Instruments Co., Ltd.) as a thermometer and heater, respectively.
We calibrated the chip resistor for its use as a thermometer
within the temperature range from 50 mK to 6 K prior to the measurements.
The stage was weakly coupled with a thermal bath (chassis of the cell made of brass)
through a thermal diffusion path.
The actual diffusion path consisted of several manganin wires with diameters of 0.025 mm (The Nilaco Corporation).
The cell was placed in the IVC of the dilution refrigerator, and thus
thermal leakage through other paths was considered negligible.

Here, we explain the measurement principles.
Assume that a constant electric power $P_0$ is applied to the heater at $t=0$ and is turned off at $t_0$ (Fig. \ref{figs3}(d)).
Correspondingly, the expected response of the sample temperature is schematically shown in Fig. \ref{figs3}(e).
From the conservation of energy, the following relationship holds at arbitrary $t$:
\begin{equation}
P(t)=C[T(t)]\dfrac{dT(t)}{dt}+\dfrac{A}{l}\int_{T_{bath}}^{T(t)}\kappa (T')dT'.
\label{eq_balance}
\end{equation}
Here, $P(t)$ is the electric power supplied by the heater and is either $P_0$ or 0.
$C$ is the combined heat capacity of the sample and addenda.
$T(t)$ is the temperature at time $t$.
$T_{bath}$ is the temperature of the thermal bath
and is monitored by a RuO$_2$ thermometer on the cell chassis.
$T_{bath}$ can be kept constant via feedback control
using a heater tightened on the cell.
$A$, $l$, and $\kappa$ are the cross-section, length, and thermal conductivity of the diffusion path, respectively.
From Eq. (\ref{eq_balance}), we can obtain the temperature dependence of $C$ as follows:
\begin{equation}
C[T(t)]=\left[P(t)-\dfrac{A}{l}\int_{T_{bath}}^{T(t)}\kappa (T')dT'\right]/\dfrac{dT(t)}{dt}.
\label{eq_sweep}
\end{equation}
If the parameters $P$, $A$, $l$, $T_{bath}$, and $\kappa (T)$ are all known,
we can obtain $C$ at arbitrary $t$ using the value of $dT(t)/dt$ obtained from the temperature response curve.
This method is called the ``sweep method''.

Here, we consider another case where the temperature increase is small enough to assume $C$ and $\kappa$ as constants [Fig. \ref{figs3}(f)].
Thus, Eq. (\ref{eq_balance}) is solved analytically and we obtain
\begin{equation}
T(t)=T_{bath}+\dfrac{P_0 l}{\kappa A}(1-\exp(-t/\tau))
\label{eq_relax_heat}
\end{equation}
for time region 1 in Fig. \ref{figs3}(f) and
\begin{equation}
T(t)=T_{bath}+\dfrac{P_0 l}{\kappa A}\exp(-t/\tau)
\label{eq_relax_cool}
\end{equation}
for time region 2 in Fig. \ref{figs3}(f).
For simplicity, we redefine $t_0$ as $t=0$ in Eq. (\ref{eq_relax_cool}).
The relaxation time $\tau=Cl/(\kappa A)$ is directly dependent on $C$,
and thus we can obtain $C$ by curve fitting procedures.
$\kappa$ contained in $\tau$ can be obtained from the temperature step $\Delta T$ as
\begin{equation}
\kappa=\dfrac{P_0 l}{\Delta T A},
\end{equation}
using the known $P_0$, $A$, and $l$.
This method is called the ``step method''.

First, we obtained the temperature dependence of $\kappa$ and addenda heat capacity $C_{addenda}$ using the step method.
The typical $\Delta T$ is $\sim1$-10 \% of $T_{bath}$.
We observed the temperature response curves at various temperatures
and obtained $\kappa(T)$ and $C_{addenda}(T)$ (Figs. \ref{figs3}(g) and \ref{figs3}(h)).
As shown in Fig. \ref{figs3}(g), $\kappa$ of manganin was observed in almost a straight line in the log-log plot, and thus we obtained the analytical expression as
$\kappa(T)=0.0757T^{1.1004}$ W/(m K) by curve-fitting.
We also confirmed that $C_{addenda}(T)$ was reasonably reproduced
by $C(T) = \alpha/T^2 + \gamma T + \beta T^3$, where
the first, second, and third terms represent nuclear, electronic, and phononic contributions, respectively.
Each coefficient obtained by curve fitting is listed in Fig. \ref{figs3}(h).

The expression of $\kappa(T)$ obtained in the above procedures allows us to evaluate the integration in Eq. (\ref{eq_sweep}). 
Compared to the step method, denser data points were obtained over the wide temperature range from a single response curve via the sweep method, and thus it was advantageous to detect a rapid change in $C$
accompanied by a phase transition.
For this reason, we adopted the sweep method in this study.

\begin{figure*}[]
\centering
\includegraphics[]{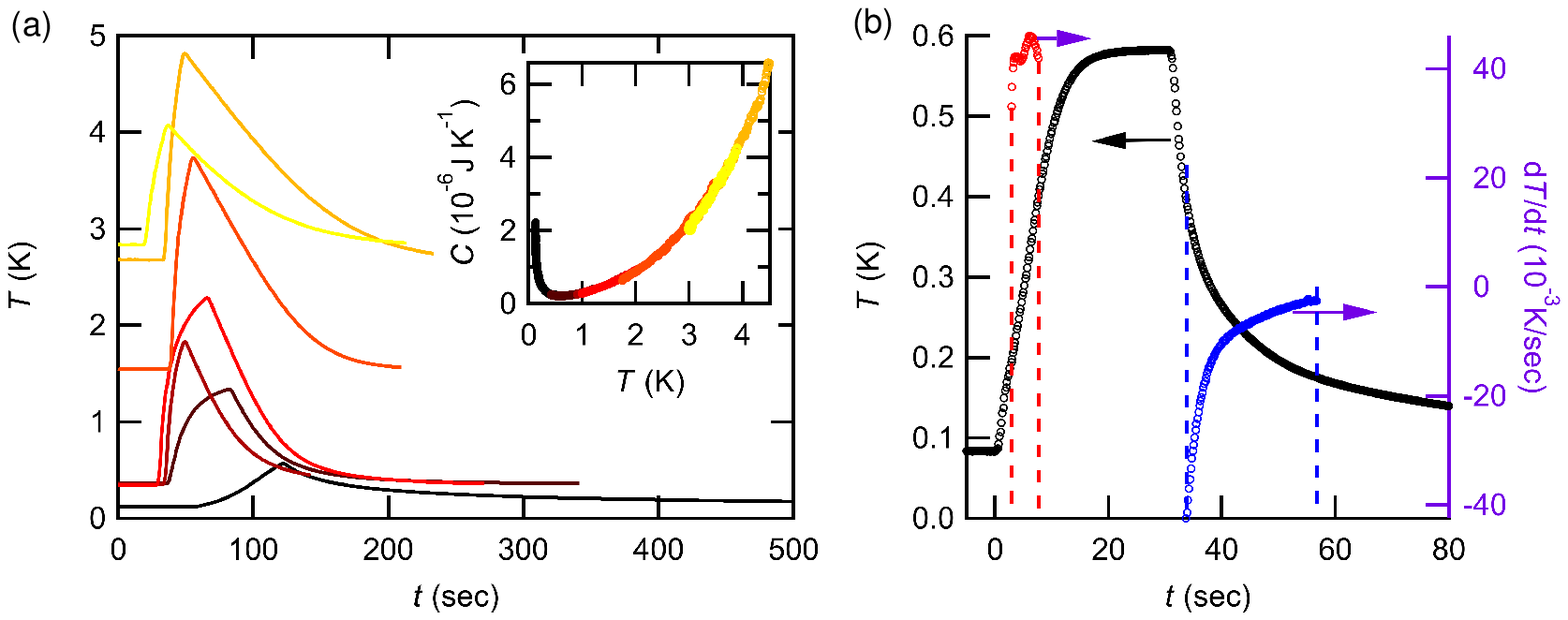}
\caption{
(a) Temperature response curves ranging from 0.1 to 5 K.
The inset shows the temperature dependence of the heat capacity obtained from the data shown in (a) using the sweep method.
(b) Temperature response curve near the superconducting transition temperature.
The right axis represents the time derivative of the response curve used in the sweep method.
\label{figs4}}
\end{figure*}

The raw data to deduce $c$ in the main text are listed in Fig. \ref{figs4}(a).
We used the cooling of each data for the analysis.
The plot of the obtained temperature dependence of $C$ is shown in the inset of Fig. \ref{figs4}(a),
in which the color corresponds to the raw response curves.
After trimming the overlapped points,
we subtracted $C_{addenda}$ from it to obtain the sample-heat capacity.
Further, it was normalized by the molar amount of the sample,
and finally, we obtained the specific heat $c$ mentioned in the main text.
The anomaly of $c$ at the superconducting transition was obtained by another measurement
using a different sample piece, and the relaxation curve is shown in Fig. \ref{figs4}(b).
To check the specific heat anomaly at $T_c^{sh}=0.27$ K,
we calculated the specific heat from the heating and cooling curves
and confirmed that the anomaly was reproducible in both processes as indicated in the main text.

\clearpage

\section{Temperature dependence of resistivity up to 300 K}

\begin{figure*}[]
\centering
\includegraphics[]{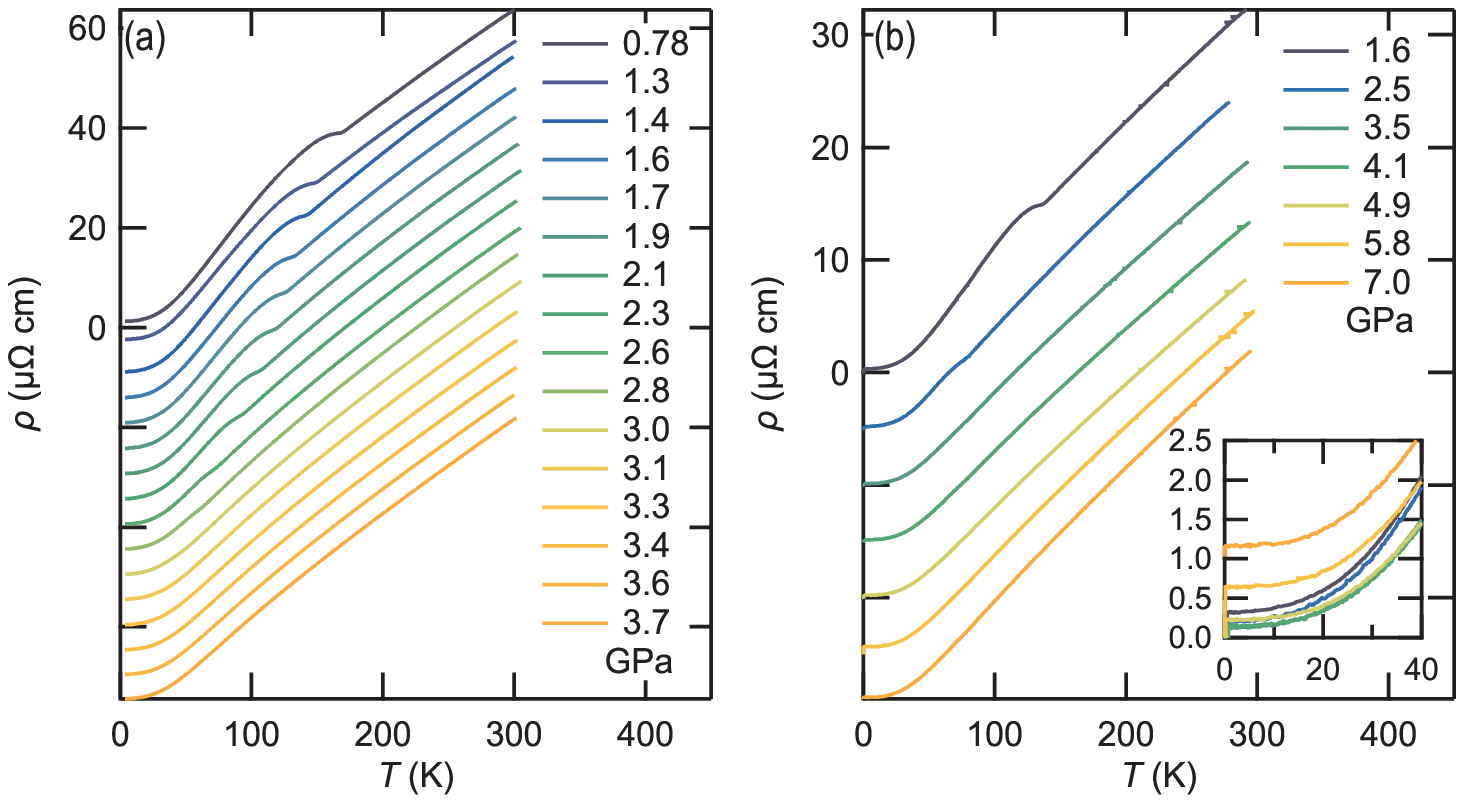}
\caption{
(a) Temperature dependence of in-plane resistivity up to 300 K at various pressures up to 3.7 GPa measured using an indenter-type pressure cell.
(b) Temperature dependence of in-plane resistivity up to 300 K at various pressures up to 7.0 GPa measured using an opposed-anvil-type pressure cell.
The data are vertically shifted for clarity.
The inset of (b) shows the magnified view below 40 K.
\label{figs5}}
\end{figure*}

Fig. \ref{figs5} shows the plot of the temperature dependence of in-plane resistivity up to 300 K for all pressures investigated here.
The data in the left and right panels were obtained using an indenter-type pressure cell and
an opposed-anvil-type pressure cell, respectively.
The inset of Fig. \ref{figs5}(b) shows the magnified view of the data obtained below 40 K.
The power $n$ of the temperature dependence was obtained by fitting the data shown in the inset with the fitting range
from 2.5 K to 35 K. 
\clearpage

\section{First-principles calculations}

\begin{figure}[]
\centering
\includegraphics[]{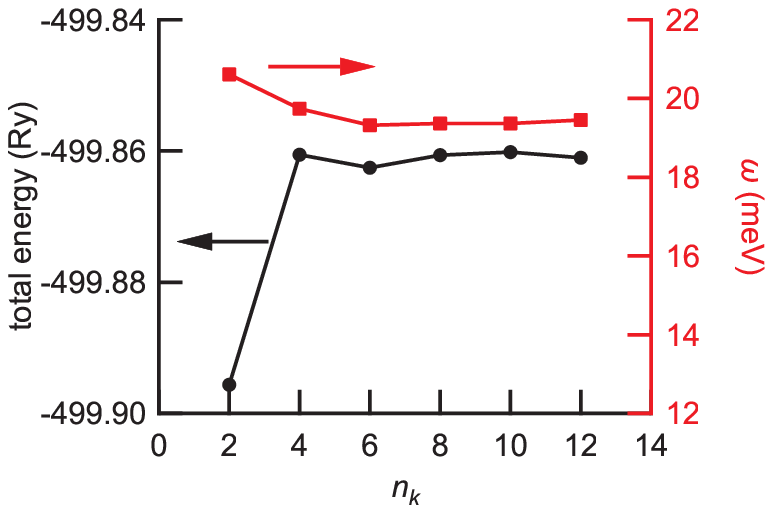}
\caption{
Total energy of the electron system (left axis) and the highest optical phonon frequency at the $\Gamma$ point (right axis) at ambient pressure calculated on the $\Gamma$-shifted $n_k^3$ Monkhorst-Pack $k$-point grids.
\label{figs6}}
\end{figure}

The electronic structure calculations and structural optimization based on the density-functional theory (DFT) were performed by the Quantum ESPRESSO (QE) package \cite{Giannozzi_2009, Giannozzi_2017}.
We used scalar-relativistic ultrasoft pseudopotentials with the Perdew-Burke-Ernzerhof exchange-correlation functional \cite{pseudo}.
We set a cut-off of 75 and 540 Ry for the plane-wave expansion of the wave functions and charge density, respectively.
Self-consistent calculations were performed with a threshold of $1.0\times 10^{-8}$ Ry.
We evaluated the total energy at ambient pressure on several $\Gamma$-shifted $n_k^3$ Monkhorst-Pack $k$-point grids
(Fig. \ref{figs6})
and adopted a $\Gamma$-shifted $6^3$ $k$-mesh in terms of its accuracy and computational efficiency for subsequent calculations.
For the structural optimization, internal atomic coordinates were relaxed in the fixed unit cell using convergence thresholds of $1.0\times 10^{-5}$ Ry for the total energy change and $1.0\times 10^{-4}$ Ry/Bohr for the forces.
We adopted experimental lattice constants \cite{Akiba_2021}
for the calculation at ambient pressure.
For calculations at 3.5 and 7.0 GPa, we used the lattice constants listed in Table \ref{tab_opt}, which were estimated from the experimental lattice compressibility \cite{Budko_2006}.
Fully-relaxed positions for La (0.25, 0.25, $z_{La}$) and Sb2 (0.75, 0.75, $z_{Sb2}$) are also shown in Table \ref{tab_opt}.
The atomic positions at ambient pressure show reasonable agreement with our experimental values $z_{La}=0.23969$
and $z_{Sb2}=0.33036$ \cite{Akiba_2021}.
From the density of states at ambient pressure,
we could estimate the electron specific heat as 2.2 mJ mol$^{-1}$ K$^{-2}$,
which was in agreement with our experimental value of 1.95 mJ mol$^{-1}$ K$^{-2}$.

\begin{table}
\caption{\label{tab_opt}
Lattice constants $a$ and $c$ used for the structural optimization and fully relaxed atomic coordinates for La (0.25, 0.25, $z_{La}$) and Sb2 (0.75, 0.75, $z_{Sb2}$).
}
\begin{ruledtabular}
\begin{tabular}{lrrrr}
$P$ (GPa) & $a$ (\AA) & $c$ (\AA) & $z_{La}$ & $z_{Sb2}$ \\
\hline
0.0 & 4.3941 & 10.868 &0.24071&0.32752\\
3.5 & 4.3495 & 10.556 &0.24154&0.32428\\
7.0 & 4.3282 & 10.379 &0.24219&0.32218
\end{tabular}
\end{ruledtabular}
\end{table}

\begin{table}
\caption{\label{tab_phfreq}
Calculated phonon frequencies at $\Gamma$ point ($\omega_\Gamma$) at ambient pressure.
$\omega_\Gamma^{ref}$ is taken from a previous calculation by Singha \textit{et al} \cite{Singha_2020}.
$\omega_\Gamma^{SO}$ is obtained by full-relativistic calculation.
``I'' and ``R'' in activity column represents infrared active and Raman active, respectively.
}
\begin{ruledtabular}
\begin{tabular}{lrrrr}
$\omega_\Gamma$ (meV) & $\omega_\Gamma^{ref}$ (meV) &$\omega_\Gamma^{SO}$ (meV) &mode & activity \\
\hline
0.00 &&0&$A_{2u}$ &I \\
0.00 &&0&$E_u$ &I \\
6.86 &4.06&6.86&$E_g$ &R \\
7.19 &&7.27&$E_u$ &I \\
9.46 &&9.43&$A_{2u}$&I  \\
10.84 &7.80&10.73&$E_g$&R  \\
11.06&&11.09&$E_u$&I  \\
12.01 &12.3&12.04&$B_{1g}$&R  \\
12.46 &13.0&12.69&$B_{1g}$&R  \\
12.55 &12.4&12.80&$E_g$&R  \\
13.02 &13.4&12.84&$A_{1g}$&R\\  
14.59 &15.1&14.43&$A_{1g}$&R\\  
15.93&&15.98&$A_{2u}$&I \\ 
16.94&&16.97&$E_u$&I  \\
17.30&&17.31&$A_{2u}$&I \\ 
19.32 &20.2&19.29&$E_g$&R  
\end{tabular}
\end{ruledtabular}
\end{table}

The phonon calculations were performed based on the density functional perturbation theory (DFPT)
with the optimized tetrahedron method \cite{Kawamura_2014} implemented in QE.
A convergence threshold of $1.0\times 10^{-14}$ Ry was employed for the DFPT self-consistent iterations.
As seen in Fig. \ref{figs6}, the highest optical phonon frequency at the $\Gamma$ point at ambient pressure
sufficiently converged on the adopted $6^3$ $k$-mesh.
The obtained frequency of 19.32 meV was consistent with the previous Raman experiment (18.7 meV) and calculation (20.2 meV) \cite{Singha_2020}.
All calculated phonon frequencies at the $\Gamma$ point at ambient pressure are listed in the $\omega_{\Gamma}$ column in Table \ref{tab_phfreq}.
Although several modes with lower frequencies exhibited relatively significant differences compared to the previous calculation \cite{Singha_2020}
($\omega_{\Gamma}^{ref}$),
the overall correspondence seemed to be reasonable.
The mismatch might be due to the difference in the utilized package and the adopted numerical condition.
The phonon dispersions were calculated using a $\Gamma$-centered $4^3$ $q$-point grid.

\begin{figure*}[]
\centering
\includegraphics[]{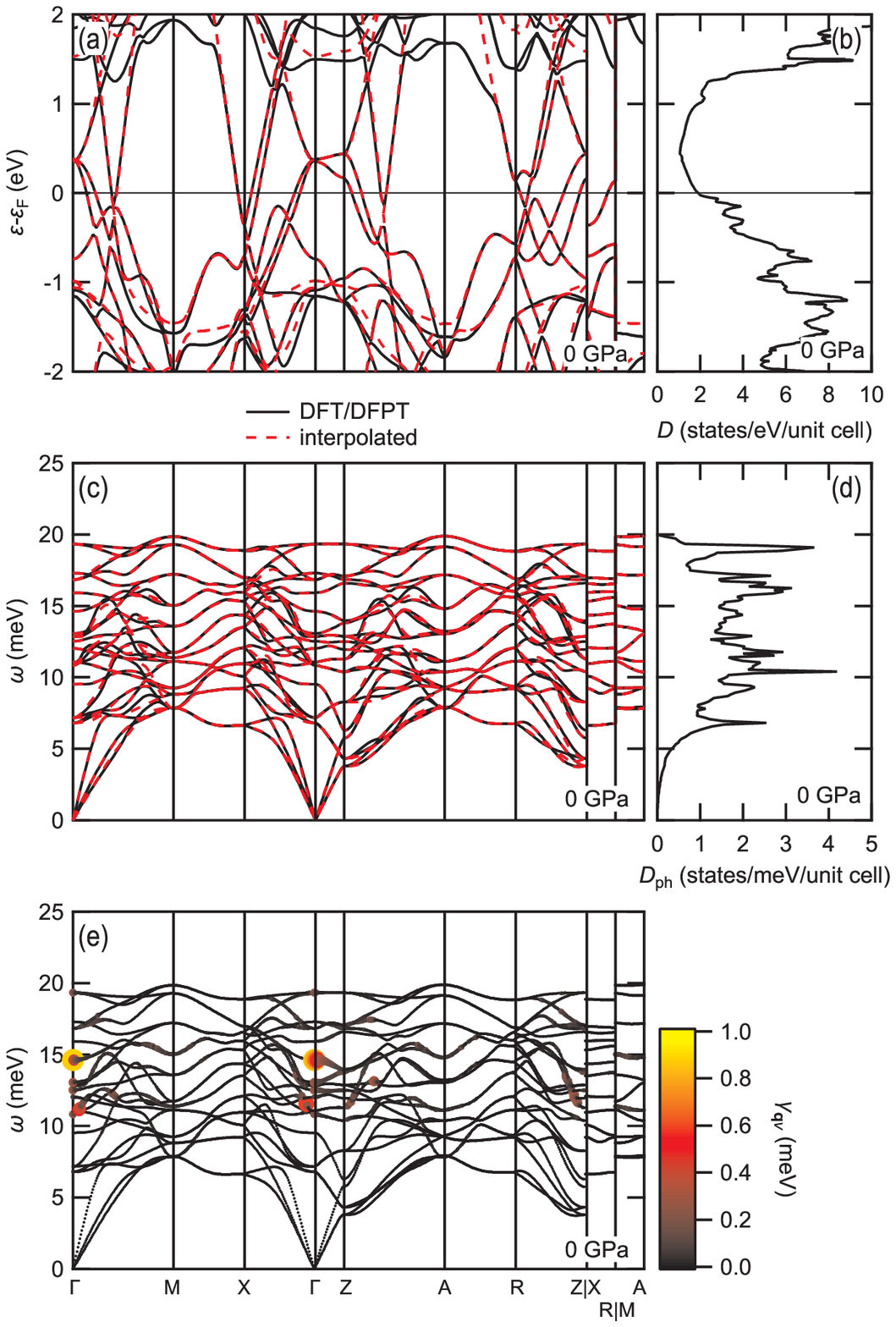}
\caption{
(a) Electron-band structure, (b) electron density of states, (c) phonon-band structure, and (d) phonon density of states at 0 GPa.
The solid black curves represent the DFT/DFPT results and red dashed curves represent the interpolated results.
(e) Distribution of the phonon linewidth $\gamma_{\bm{q}\nu}$ at 0 GPa.
The color and size of the markers represent the magnitude of $\gamma_{\bm{q}\nu}$.
\label{figs7}}
\end{figure*}

\begin{figure*}[]
\centering
\includegraphics[]{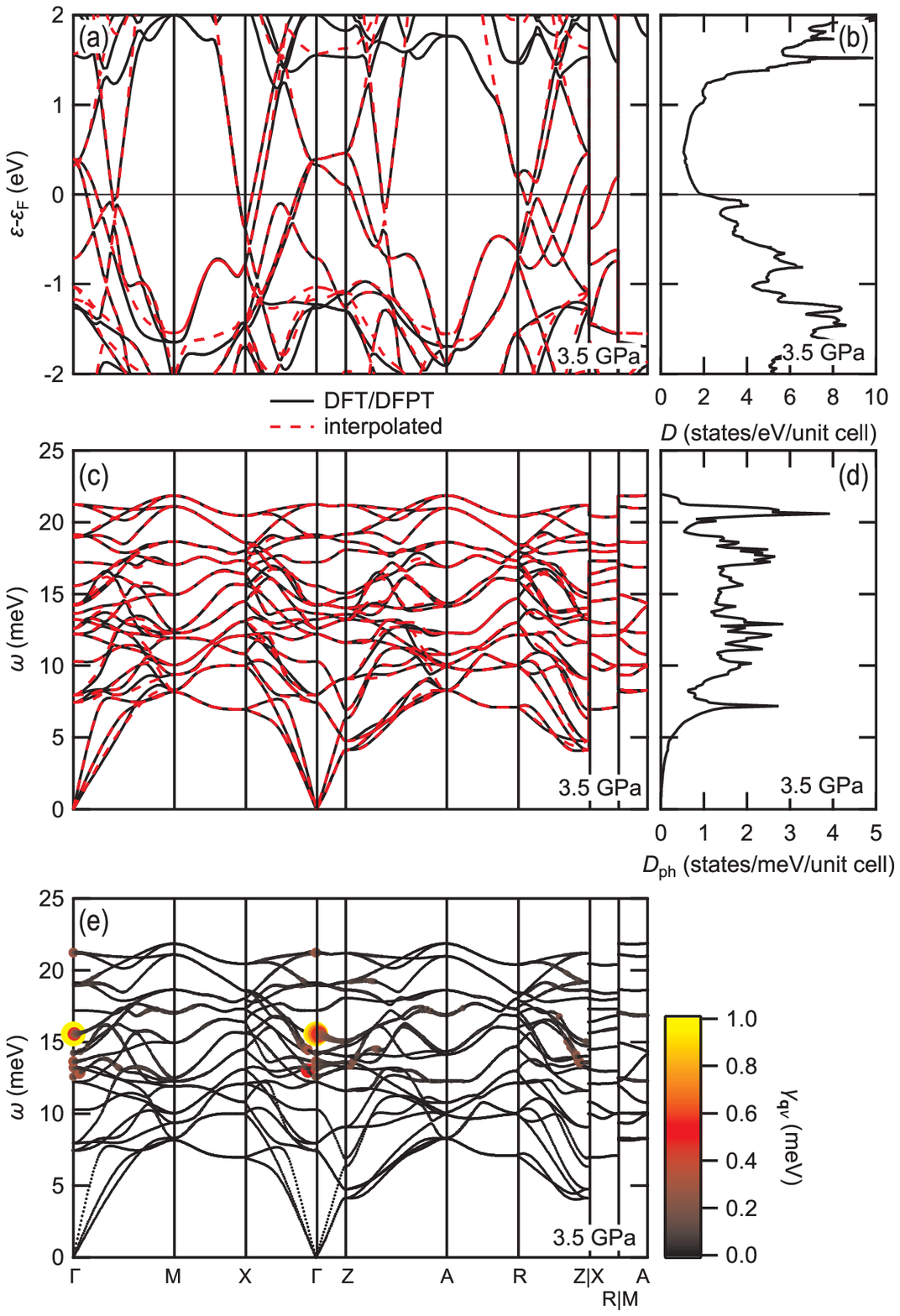}
\caption{
(a) Electron-band structure, (b) electron density of states, (c) phonon-band structure, and (d) phonon density of states at 3.5 GPa.
The solid black curves represent the DFT/DFPT results and red dashed curves represent the interpolated results.
(e) Distribution of the phonon linewidth $\gamma_{\bm{q}\nu}$ at 3.5 GPa.
The color and size of the markers represent the magnitude of $\gamma_{\bm{q}\nu}$.
\label{figs8}}
\end{figure*}

\begin{figure*}[]
\centering
\includegraphics[]{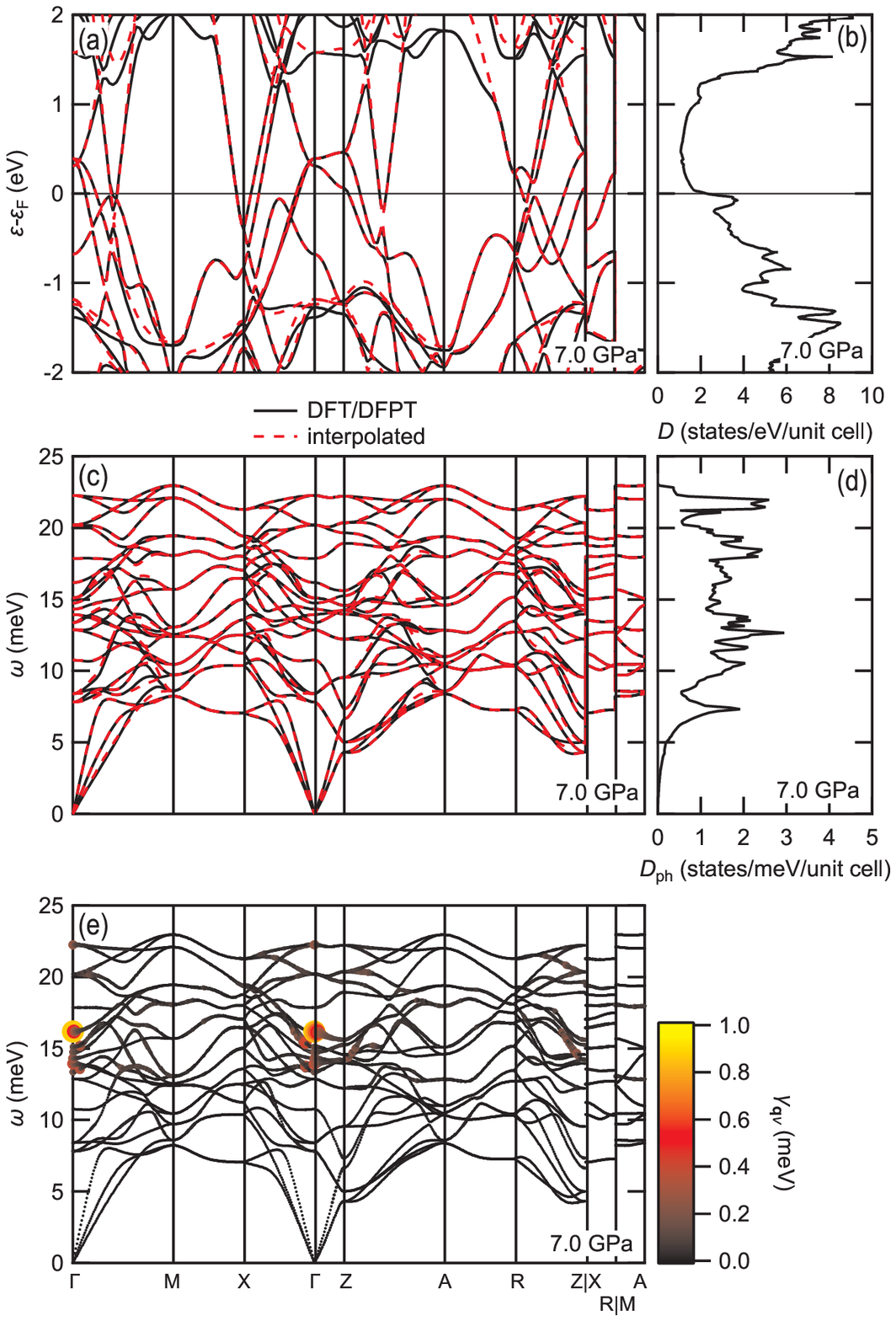}
\caption{
(a) Electron-band structure, (b) electron density of states, (c) phonon-band structure, and (d) phonon density of states at 7.0 GPa.
The solid black solid represent the DFT/DFPT results and red dashed curves represent the interpolated results.
(e) Distribution of the phonon linewidth $\gamma_{\bm{q}\nu}$ at 7.0 GPa.
The color and size of the markers represent the magnitude of $\gamma_{\bm{q}\nu}$.
\label{figs9}}
\end{figure*}

We plotted the (a) electron-band structure $\epsilon-\epsilon_F$,
(b) electron density of states $D(\epsilon)$,
(c) phonon-band structure $\omega$,
and (d) phonon density of states $D_{ph}(\omega)$ obtained by DFT/DFPT calculations
in Figs. \ref{figs7}--\ref{figs9} (black solid curves).

Here, we comment on the effect of spin-orbit coupling (SOC).
In this study, we adopted scalar-relativistic calculations without SOC.
In previous studies, it has already been shown that SOC causes only a slight change in the electronic band structure
around the Fermi level \cite{Ruszala_2020, Akiba_2022}.
To test the effect of SOC on phonon properties, we calculated the phonon frequency at the $\Gamma$ point
using full-relativistic ultrasoft pseudopotentials, and its results are shown in the $\omega_\Gamma^{SO}$ column
in Table \ref{tab_phfreq}.
Compared with the scalar-relativistic case, the difference is only less than three percent for all modes, and thus we can adopt the scalar-relativistic calculation here.

\begin{figure}[]
\centering
\includegraphics[]{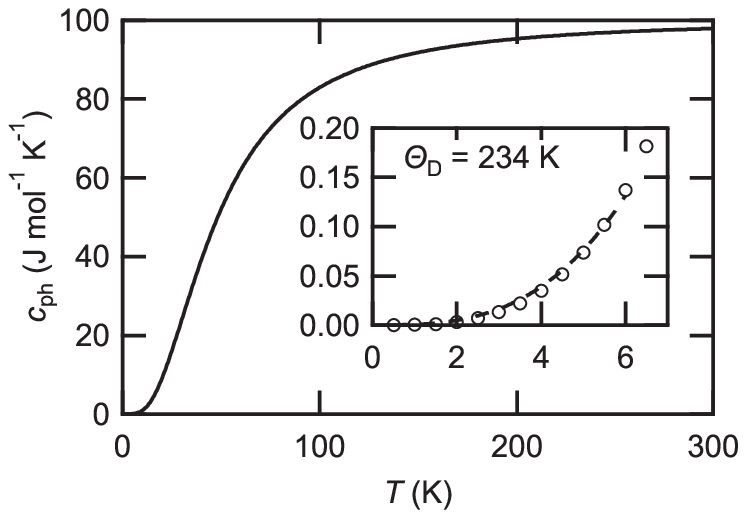}
\caption{
Temperature dependence of the phonon specific heat calculated
from the theoretical phonon density of states using Eq. (\ref{eq_cph_calc}).
The inset shows the magnified view of the low temperature region with fitting curves (dashed curve) assuming that $c_{ph}\propto T^3$.
\label{figs10}}
\end{figure}

From the theoretical phonon density of the state spectrum $D_{ph}(\omega)$,
we can obtain the temperature dependence of the phonon specific heat $c_{ph}(T)$ as \cite{Ziman_ep}
\begin{equation}
c_{ph}(T)=\int d\omega k_B\left(\dfrac{\hbar \omega}{k_B T}\right)^2 \dfrac{\exp\left(\dfrac{\hbar \omega}{k_B T}\right)}{\left[\exp\left(\dfrac{\hbar \omega}{k_B T}\right)-1\right]^2}D_{ph}(\omega),
\label{eq_cph_calc}
\end{equation}
and the calculated $c_{ph}(T)$ at ambient pressure is represented in Fig. \ref{figs10}.
We can evaluate the Debye temperature as 234 K by curve fitting at low temperature up to 6 K (inset of Fig. \ref{figs10}),
which is consistent with our experimental value of 260 K.
We also confirmed quite satisfactory agreement with the previous
heat capacity measurement up to 80 K \cite{Takeuchi_2003}.
The above consistencies with previously reported experiments confirm the validity of the calculated phonon properties.

\clearpage

\section{Electron-phonon coupling calculations}

Based on the DFT and DFPT results, we calculated the electron-phonon coupling
and phonon-mediated superconducting transition temperature using Wannier90 \cite{Pizzi_2020} and EPW \cite{Ponce_2016} codes.
The electron-phonon matrix element is defined as 
\begin{equation}
g_{mn, \nu}(\bm{k}, \bm{q})=\sqrt{\dfrac{\hbar}{2M \omega_{\bm{q}\nu}}}\langle \Psi_{m\bm{k}+\bm{q}}|\partial_{\bm{q}\nu}V|\Psi_{n\bm{k}}\rangle.
\end{equation}
Here, $M$ and $\hbar$ are the mass of the nuclei and reduced Planck constant, respectively.
$\omega_{\bm{q}\nu}$ represents the frequency of phonon with a wavevector $\bm{q}$ and mode index $\nu$. 
$|\Psi_{n\bm{k}}\rangle$ is the electronic wavefunction for band index $n$ and wavevector $\bm{k}$
with eigenvalue of $\epsilon_{n\bm{k}}$.
$\partial_{\bm{q}\nu}V$ is the derivative of the self-consistent potential associated with a phonon having $\bm{q}$ and $\nu$.
Using $g_{mn, \nu}(\bm{k}, \bm{q})$, 
the phonon linewidth $\gamma_{\bm{q}\nu}$ and electron-phonon coupling strength $\lambda_{\bm{q}\nu}$ associated with a phonon having $\nu$ and $\bm{q}$ are represented as
\begin{equation}
\gamma_{\bm{q}\nu}=2\pi \omega_{\bm{q}\nu}\sum_{nm}\int_{BZ}\dfrac{d\bm{k}}{\Omega_{BZ}} |g_{mn, \nu}(\bm{k}, \bm{q})|^2\delta(\epsilon_{n\bm{k}}-\epsilon_F)\delta(\epsilon_{m\bm{k}+\bm{q}}-\epsilon_F),
\label{eq_linewidth_int}
\end{equation}
\begin{equation}
\lambda_{\bm{q}\nu} = \dfrac{2}{N(\epsilon_F)\omega_{\bm{q}\nu}} \sum_{nm}\int_{BZ}\dfrac{d\bm{k}}{\Omega_{BZ}} |g_{mn, \nu}(\bm{k}, \bm{q})|^2\delta(\epsilon_{n\bm{k}}-\epsilon_F)\delta(\epsilon_{m\bm{k}+\bm{q}}-\epsilon_F)=\dfrac{\gamma_{\bm{q}\nu}}{\pi N(\epsilon_F)\omega_{\bm{q}\nu}^2}.
\label{eq_lambda_int}
\end{equation}
Here, $N(\epsilon_F)$ is the density of states per spin at the Fermi level $\epsilon_F$,
and the integral is taken over the Brillouin zone with $\Omega_{BZ}$ volume.
$\delta(\epsilon)$ represents the Dirac delta function.
Alternatively, we can represent the electron-phonon coupling strength
in the $\bm{k}$-space as
\begin{equation}
\lambda_{n\bm{k}}=\sum_{m\nu}\int_{BZ}\dfrac{d\bm{q}}{\Omega_{BZ}}\dfrac{2}{\omega_{\bm{q}\nu}}|g_{mn, \nu}(\bm{k}, \bm{q})|^2\delta(\epsilon_{m\bm{k}+\bm{q}}-\epsilon_F).
\end{equation}
In the main text and hereafter, we omit the band index $n$ and simply write as $\lambda_{\bm{k}}$.

Using $\lambda_{\bm{q}\nu}$, the Eliashberg spectral function $\alpha^2F(\omega)$ can be obtained by calculating its integrated value over the Brillouin zone as follows:
\begin{equation}
\alpha^2F(\omega) = \dfrac{1}{2}\sum_{\nu}\int_{BZ}\dfrac{d\bm{q}}{\Omega_{BZ}} \omega_{\bm{q}\nu} \lambda_{\bm{q}\nu} \delta(\omega-\omega_{\bm{q}\nu}).
\label{eq_a2f_int}
\end{equation}
Then, we estimated the superconducting transition temperature $T_c^{MAD}$ using the McMillan-Allen-Dynes formula \cite{McMillan_1968, Dynes_1972, Allen_1975}
\begin{equation}
T_c^{MAD} = \dfrac{\omega_{log}}{1.2}\exp \left(-\dfrac{1.04(1+\lambda)}{\lambda-\mu_c^*(1+0.62\lambda)}\right).
\end{equation}
Here, $\lambda$ is defined using the Eliashberg spectral function and maximum phonon frequency $\omega_{max}$ as
\begin{equation}
\lambda = 2\int_0^{\omega_{max}} d\omega \dfrac{\alpha^2F(\omega)}{\omega},
\end{equation}
and $\omega_{log}$ is a logarithmic average of the phonon frequency defined as
\begin{equation}
\omega_{log}=\exp\left( \dfrac{2}{\lambda}\int_0^{\omega_{max}} d\omega \ln\omega \dfrac{\alpha^2F(\omega)}{\omega}\right).
\end{equation}
$\mu_c^*$ represents the Coulomb pseudopotential, which is treated as an empirical parameter.
For typical metals, $\mu_c^*$ is known to take values around 0.1 \cite{Morel_1962}.
In the main test, we assumed $\mu_c^*=0.1$.

\begin{figure*}[]
\centering
\includegraphics[]{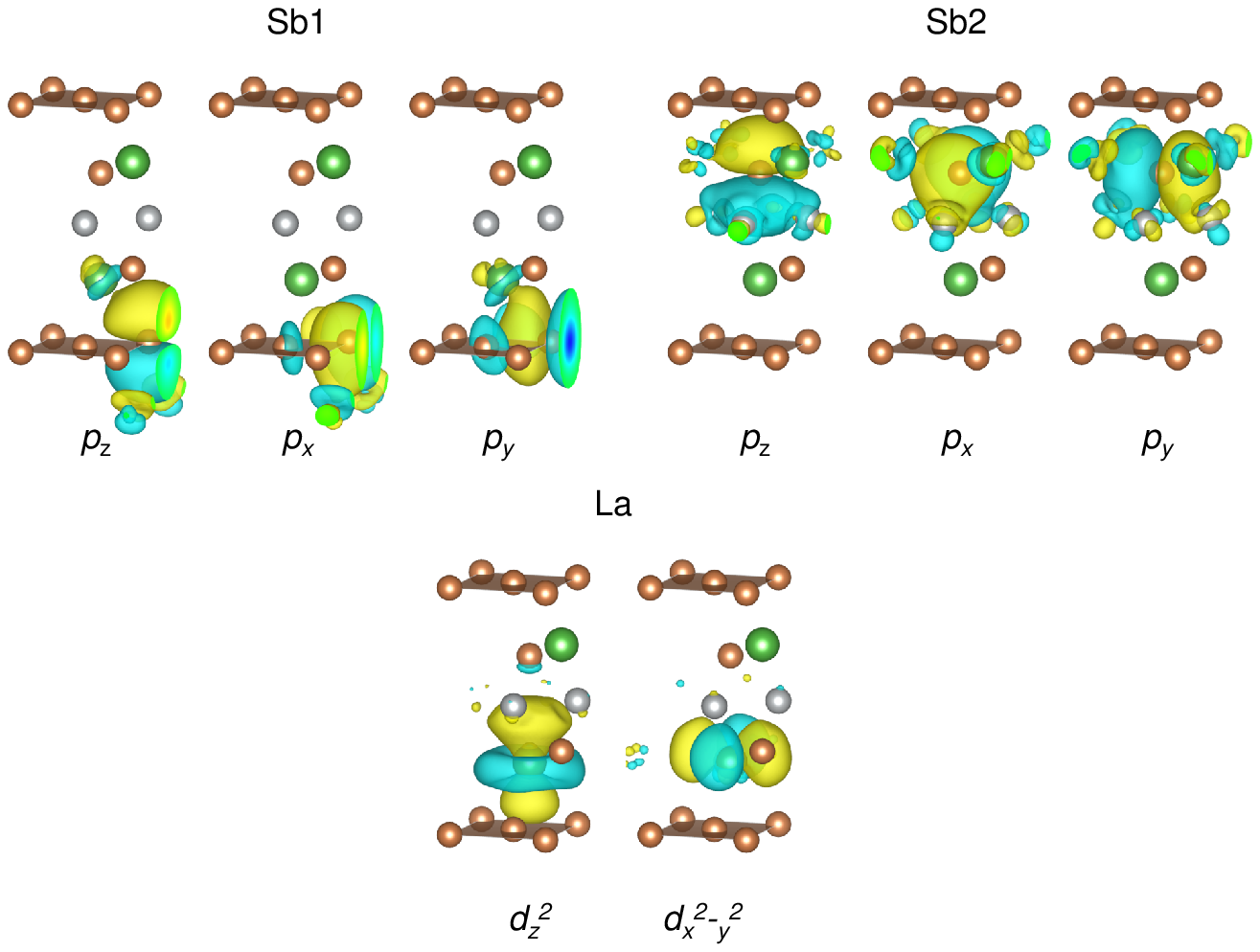}
\caption{
Obtained maximally-localized Wannier orbitals at ambient pressure.
\label{figs11}}
\end{figure*}

We used coarse $8^3$ $k$-mesh and $4^3$ $q$-mesh
for the initial calculation of the electronic Hamiltonian, dynamical, and electron-phonon matrices.
To calculate the electron-phonon coupling properties on arbitrary dense Brillouin zone grids,
an interpolation scheme described in \cite{Giustino_2007} was applied using EPW.
In this procedure, we used 16 maximally localized Wannier functions.
We show in Figs. \ref{figs7}--\ref{figs9}(a--d) the comparison of DFT/DFPT band structures
(black solid curves) and interpolated band structures (red dashed curves)
at 0, 3.5, and 7.0 GPa,
indicating that the interpolation procedure works well with sufficient accuracy.
The resulting maximally-localized Wannier orbitals at ambient pressure are shown in Fig. \ref{figs11}.

\begin{figure}[]
\centering
\includegraphics[]{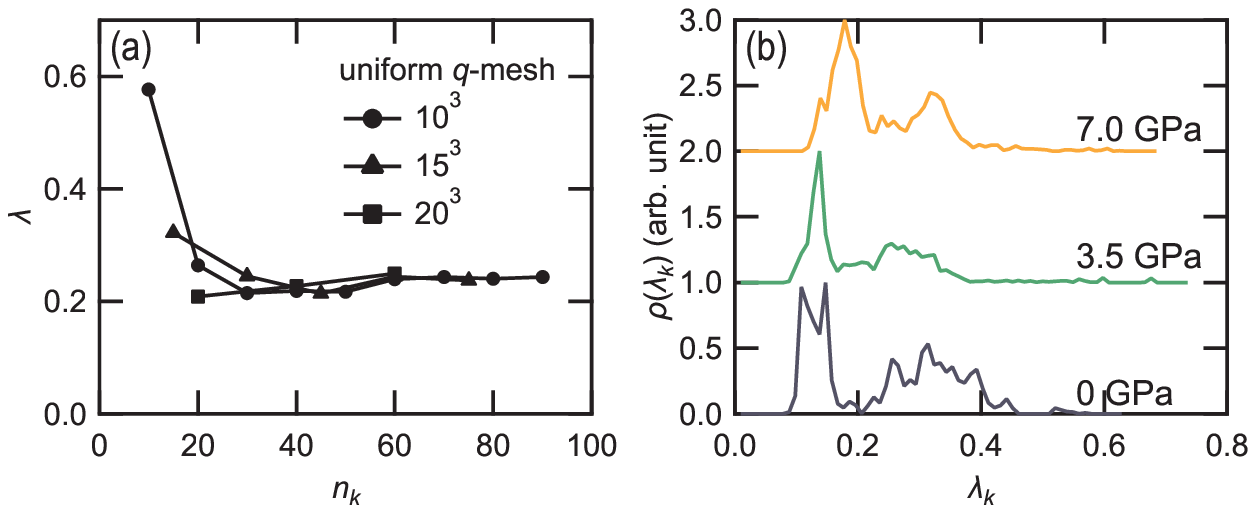}
\caption{
(a) Electron-phonon coupling strength $\lambda$ at ambient pressure calculated for
various sampling conditions.
We used uniform $n_k^3$ $k$-meshes.
(b) Population ($\rho(\lambda_{\bm{k}})$) of the electron-phonon coupling strength ($\lambda_{\bm{k}}$) on the Fermi surface.
We considered the carriers within the energy range of $\pm 0.2$ eV from the Fermi energy.
\label{figs12}}
\end{figure}

The integrations over the Brillouin zone were performed 
on uniform $75^3$ $k$-mesh and $15^3$ $q$-mesh \cite{conv}.
The Dirac delta functions
were smeared with widths of 25 meV for electrons
and 0.05 meV for phonons.
As shown in Fig. \ref{figs12}(a), the adopted meshes were sufficient to obtain the converged $\lambda$
and were optimal with regard to the computational costs.
The distribution of the phonon linewidth $\gamma_{\bm{q}\nu}$ at each pressure is indicated in Figs. \ref{figs7}--\ref{figs9}(e).
By integrating $\gamma_{\bm{q}\nu}$ over the Brillouin zone according to Eqs. (\ref{eq_lambda_int}) and (\ref{eq_a2f_int}),
we can obtain the Eliashberg spectral function $\alpha^2F(\omega)$.

Finally, we show in Fig. \ref{figs12}(b) the population of the electron-phonon coupling strength $\lambda_{\bm{k}}$ on the Fermi surface.
Here, we consider the carriers within the energy range of $\pm 0.2$ eV from the Fermi energy.
We can recognize two distinct peaks at all pressures.
The secondary peak at approximately $\lambda_{\bm{k}}\sim 0.3$ originates from the carriers on the hollow-like Fermi surface.

\clearpage

\section{Orbital character of the Fermi surface}
\begin{figure*}[]
\centering
\includegraphics[]{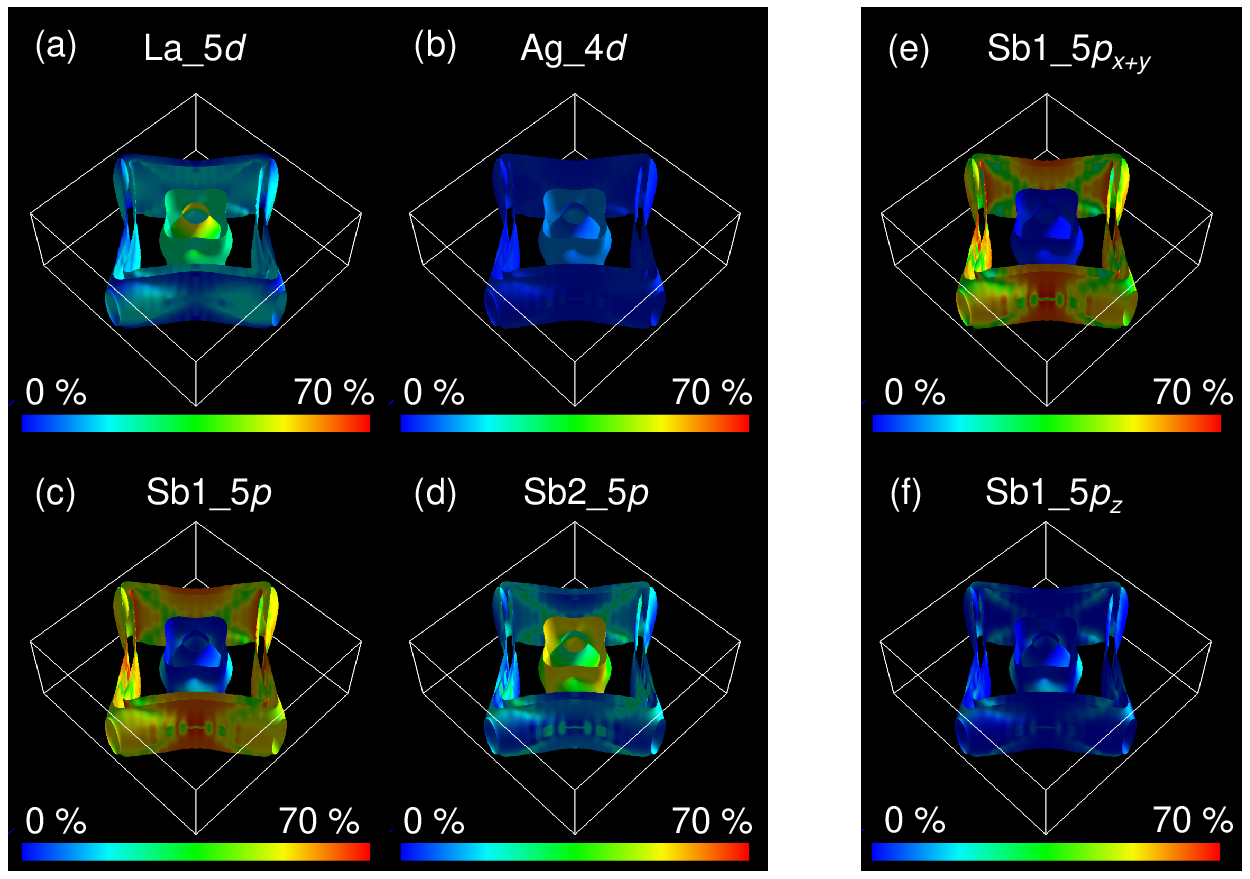}
\caption{
Projection of (a) La-5$d$, (b) Ag-4$d$, (c) Sb1-5$p$, and (d) Sb2-5$p$ characters on the Fermi surface at ambient pressure.
Separate views of (e) Sb1-5$p_{x+y}$ and (f) Sb1-5$p_{z}$ characters.
\label{figs13}}
\end{figure*}

In Fig. \ref{figs13}, we show the projections of atomic orbitals $|\langle \Psi_{atom, orbital}|\Psi_{n\bm{k}}\rangle|^2$
on the Fermi surface at ambient pressure.
The calculation was obtained on the $36^3$ $k$-point grid and visualized using FermiSurfer \cite{Kawamura_2019}.
We plot the projections of representative orbitals having major density of states at the Fermi level,
i.e., (a) La-5$d$, (b) Ag-4$d$, (c) Sb1-5$p$, and (d) Sb2-5$p$.
Here, Sb1 constitutes the square net layer.
We also show in Figs. \ref{figs13}(e) and (f) the details of Sb1-5$p$ orbital.
The hollow-like Fermi surface is mostly constructed by 5$p_{x+y}$,
and the contribution of 5$p_{z}$ is relatively minor. 


\bibliography{reference_sm}